\documentclass[journal]{IEEEtran}

\usepackage{cite}
\usepackage{amssymb}
\usepackage{graphicx}

\usepackage[cmex10]{amsmath}
\usepackage{mdwmath}
\usepackage{verbatim,textcomp}
\usepackage{circledtext}
\usepackage[table]{xcolor}
\usepackage[lined,boxed,commentsnumbered, ruled]{algorithm2e} 
\usepackage{amsthm} 

\usepackage{chemarrow}
\usepackage{extarrows}
\usepackage{multirow}  
\usepackage{array} 
\usepackage{subfigure} 
\usepackage{stfloats}
\usepackage{color}
\usepackage[colorlinks,
            linkcolor=black,
            anchorcolor=black,
            urlcolor=black,
            citecolor=black]{hyperref}
\usepackage[hyphenbreaks]{breakurl}
\allowdisplaybreaks[4]
\newtheorem{theorem}{Theorem}

\newcommand{\systemname}{{{SAGIN}}}
\newcommand{\leoname}{LEO}

\newcommand{\methodname}{DRL-and-Perception-aided Approach}
\begin{document}
\title{Cost-Efficient Computation Offloading in SAGIN: A Deep Reinforcement Learning and Perception-Aided Approach}
\author{Yulan Gao,\IEEEmembership{~Member, IEEE,}
Ziqiang Ye, \IEEEmembership{}
and Han Yu,\IEEEmembership{~Senior Member, IEEE}
\thanks{Y. Gao and H. Yu are with the School of Computer Science and Engineering, Nanyang Technological University, 639798 Singapore (e-mail: yulangaomath@163.com, han.yu@ntu.edu.sg).}
\thanks{ Z. Ye is  with the National Key Laboratory of Wireless Communications, University of Electronic Science and Technology of China (UESTC), Chengdu 611731, China (e-mail: yysxiaoyu@hotmail.com).}
}

\markboth{~}
{Shell \MakeLowercase{\textit{et al.}}: }

\maketitle
\begin{abstract}
The Space-Air-Ground Integrated Network (\systemname{}), crucial to the advancement of sixth-generation (6G) technology, plays a key role in ensuring universal connectivity, particularly by addressing the communication needs of remote areas lacking cellular network infrastructure.
This paper delves into the role of unmanned aerial vehicles (UAVs) within \systemname{}, where they act as a control layer owing to their adaptable deployment capabilities and their intermediary role. 
Equipped with millimeter-wave (mmWave) radar and vision sensors, these UAVs are capable of acquiring multi-source data, which helps to diminish uncertainty and enhance the accuracy of decision-making.  
Concurrently, UAVs collect tasks requiring computing resources from their coverage areas, originating from a variety of mobile devices moving at different speeds. 
These tasks are then allocated to ground base stations (BSs), low-earth-orbit (\leoname{}) satellite, and local processing units to improve processing efficiency. 
Amidst this framework, our study concentrates on devising dynamic strategies for facilitating task hosting between mobile devices and UAVs, offloading computations, managing associations between UAVs and BSs, and allocating computing resources.
The objective is to minimize the time-averaged network cost, considering the uncertainty of device locations, speeds, and even types.
To tackle these complexities, we propose a deep reinforcement learning and perception-aided online approach (\methodname{}) for this joint optimization in \systemname{}, tailored for an environment filled with uncertainties.
The effectiveness of our proposed approach is validated through extensive numerical simulations, which quantify its performance relative to various network parameters.
\end{abstract}
\IEEEpeerreviewmaketitle

\section{Introduction}
\IEEEPARstart{A}{dvances} in wireless communication technologies and the miniaturization of electronic components have driven the widespread adoption of smart devices, including wearables, smartphones, home appliances, and environmental monitoring sensors, across both urban and remote areas \cite{agiwal2016next}.
These applications require not only broad network coverage but also the capability to store and process large amounts of real-time data.
This situation places considerable strain on network infrastructures and service provisions, significantly increasing the demands on these systems. 
Devices involved often face limitations due to their low power and computing capabilities, yet they are tasked with executing computation-intensive operations, which presents significant  challenges \cite{tang2021deep}.  
Furthermore, the availability of fifth-generation and beyond (5G and B5G) networks is virtually absent in remote areas \cite{cavalcante20215g}. 
This lack of coverage complicates the ability of devices in these areas to establish connections through current stand-alone terrestrial networks, becoming a major hurdle in the efficient acquisition and transmission of real-time data \cite{smirnov2023real}.
By leveraging the complementary strengths of space, air, and ground network segments, Space-Air-Ground Integrated Networks (\systemname{}), along with Mobile Edge Computing (MEC), offer effective solutions for achieving high-quality, ubiquitous communications and meeting computing demand \cite{liu2023online,nguyen2022computation}.

\systemname{} embodies a sophisticated, multi-layered network architecture, intricately woven through three distinct yet interconnected segments: space, air, and ground. 
The space component of this network is categorized by altitude into geostationary-earth-orbit (GEO), medium-earth-orbit (MEO), and low-earth-orbit (LEO) satellites. Each category serves distinct purposes, offering always-on and high-capacity cloud computing services along with varying coverage areas.
In the aerial domain, networks are divided by altitude into high altitude platforms (HAPs) and low altitude platforms (LAPs), covering various vehicles like unmanned aerial vehicles (UAVs), balloons, airplanes, and airships \cite{gonzalo2018capabilities,klemas2015coastal}. 
These components are crucial for closing connectivity gaps in areas with scarce terrestrial infrastructure. 
They serve as on-demand relays or edge processors to support ground communications.
The quality of the air-to-ground communication channel is more likely to be better due to direct line-of-sight communications, which significantly reduces the power and energy requirements of ground-based devices, thereby extending their service life. 
Furthermore, deploying communication systems in the aerial layer offers both flexibility and cost efficiency, making it an advantageous strategy for expanding network reach and capacity.
On the terrestrial front, the ground network encompasses established communication infrastructures like cellular networks and wireless local area networks (WLAN). This layer forms the foundation of traditional connectivity, supporting the seamless integration with the aerial and space segments to enhance overall network performance and reliability, especially in challenging environments where conventional infrastructure may fall short.

Although existing studies 
(\cite{wang2017taking,chen2021energy,liu2022energy,liao2021learning}) have extensively examined the performance of merging MEC with \systemname{}, there are still several open issues.
First, the majority of these studies have concentrated on the short-term performance of \systemname{} by joint task offloading and resource allocation.
Yet, the actual dynamics of device task arrivals, transmissions, and processing in \systemname{} exhibit random fluctuations over time. 
Additionally, there is uncertainty in the behavior of users and the environment.
This randomness becomes particularly problematic when ground devices move rapidly, potentially leading to task execution failures. 
Therefore, there is a crucial need for designing dynamic computation offloading and resource scheduling strategies that focus on the long-term performance, aiming to boost both success rates and sustainability. 
Additionally, some studies \cite{qiu2019deep,zhou2020deep,liu2021joint} have sought to optimize long-term network performance through resource scheduling across various layers, demonstrating the importance of strategic long-term planning in enhancing network efficiency and reliability.
However, these studies are conducted based on the assumption that the channel information from air to ground is perfectly known and that the ground devices remain stationary.
While this assumption helps to create theoretical models and initial frameworks for understanding \systemname{}, practical implementations need to consider the inherent unpredictability and dynamics of real-world scenarios.

To bridge these important gaps, in this paper, we consider a scenario where a large number of ground devices that move at different speeds as they travel between urban areas, which have cellular network coverage, and remote areas lacking such coverage. 
Within the \systemname{} architecture, UAVs are outfitted with millimeter-wave (mmWave) radar and vision sensors.
Visual sensors are deployed to capture real-time imagery, which is then processed by the {\tt YOLOv7} algorithm \cite{wang2023designing,wang2023yolov7}, specifically designed to enhance the detection capabilities of the accompanying radar system.
This equips the UAVs with the capability to gather multi-source data, thereby reducing uncertainty and enhancing the accuracy of decision-making processes.
Additionally, these UAVs are tasked with gathering computing-intensive tasks from within their coverage areas, originating from various mobile devices moving at different velocities. 
These tasks are subsequently offloaded to nearby base stations (BSs), LEOs, and local processing units, aiming to minimize the time-averaged network cost and ensure the stability of all task queues.
To tackle the challenges posed by the undertain distribution of user behaviors, as well as the time-varibale and complex environment of \systemname{}, we propose a deep reinforcement learning and perception-aided approach (\methodname{}). 
This approach is designed to optimize task hosting, computation offloading, and the control of associations between UAVs and BSs, as well as the allocation of computational resources between BSs and UAVs. 
The objective is to adapt effectively to the aforementioned dynamics, striving to establish a sustainable and cost-effective SAGIN network.
The novelty and contributions of this paper are summarized in the following aspects.
\begin{itemize}
\item  We propose a novel framework that integrates visual recognition with mmWave radar detection into the joint optimization process in \systemname{}.
Specifically, when subjects are initially detected by the mmWave radar, visual capture devices are activated to further classify the detected objects and recognize their behaviors.
This integration of visual recognition and radar detection markedly enhances the decision-making process. It ensures that the combined strategies for task offloading, resource allocation, and association control are finely tuned, effectively accommodating the complexities of an environment marked by uncertain entity behaviors and a dynamically changing user base.

\item To tackle the problem of minimizing time-averaged network costs while ensuring network stability, we initially transform the problem, which is inherently linked over time, into a sequential online joint optimization problem (JOP) by Lyapunov drift-plus-penalty algorithm.
Subsequently, the JOP is segmented into three subproblems, each resolved independently using the Deep Deterministic Policy Gradient (DDPG) and the self-adaptive global best harmony search (SGHS) algorithms.

\item  Numerical simulations are used to validate the analysis and evaluate the performance of the proposed \methodname{} as a function of the \systemname{} parameters. 
Detailed results are first presented for the convergence and effectiveness of the proposed algorithms, specifically DDPG and SGHS, which address subproblems {\bf P1} and {\bf P3}. Following this, the effectiveness of the complete \methodname{} process is thoroughly evaluated.
To gain insight into the proposed \methodname{}, subsequent analysis focuses on the performance of both UAV and BS queue backlogs, noting that shorther queue backlogs indicate greater network stability. 
Further simulation results demonstrate that \methodname{} not only achieves the lowest system operation cost but also ensures network stability.
\end{itemize}

The rest of the paper is organized as follows. 
The related works are briefly summarized in Section II. 
System model and computation offloading model are presented in Section III and IV, respectively. 
Section V present the problem formulation and algorithm design. 
Section VI reports the simulation setup and results analysis.
Finally, we conclude this paper in Section VII.

\section{Related Works}
\subsection{Computation Offloading and Resource Allocation in \systemname{}}
In the landscape of computation offloading and resource allocation, \systemname{} stands as a complex framework that incorporates various critical aspects.  
These dimensions include hybrid edge-cloud computing, subject to maximum delay constraints, the intricacies of multi-hop satellite communications, optimization of UAV trajectories, the intricacies of parallel offloading control, as well as sophisticated strategies for user association-scheduling and rigorous admission control mechanisms.

The research community has delved into various dimensions of this field. 
For instance, \cite{chen2021energy} focused on jointly optimizing task scheduling and computation resource allocation, aiming to minimize the expected latency within a network featuring multiple satellites and UAVs with predefined flight trajectories.
Moreover, the broader realm of computation offloading, which includes parallel task processing across mobile devices, MEC, and cloud servers, has been thoroughly investigated in works by \cite{mao2020joint,feng2021joint,liu2022user}.
Specifically, \cite{mao2020joint} targeted the minimization of the maximum delay encountered by mobile devices by co-optimizing various elements, including UAV-device association and power and bandwidth allocation, alongside computation resources and UAV placement.
\cite{feng2021joint} embarked on reducing the average latency within a multi-user MEC setup, emphasizing the optimization of user association and task segmentation for tasks that are both independent and interconnected. 
\cite{liu2022user} considered a \systemname{} scenario with a single satellite, UAV, and multiple small cells, with the goal of maximizing the sum rate through the collective optimization of user connections, sub-channel assignments, and power distribution, constrained by maximum delay constraint.
More recently, \cite{huang2024joint} presented a novel deep reinforcement learning (DRL)-based framework for optimizing task offloading and resource allocation in a hybrid cloud and multi-access edge computing environment within SAGINs. By incorporating multiple satellites, clouds, and unmanned aerial vehicles, the proposed system effectively reduces energy consumption and latency, demonstrating superior performance and potential for practical applications in dynamic and complex communication scenarios.
\cite{jia2024service} presented a Service-Oriented SAGIN with Pervasive Intelligence, focusing on resource-constrained users within 6G communication networks. It introduced a system architecture that integrates Edge Intelligence (EI) by combining AI with MEC, enhancing the capabilities for communication, computing, sensing, and storage. The research proposes a DRL-based algorithm for resource allocation and computation offloading, demonstrating significant system cost reductions compared to existing methods, and efficiently managing resources in dynamic service environments.

\subsection{Perception and Communication Integration}
The existing literature on traditional sensing technologies is vast, covering a wide range of fields including radar sensing \cite{wang2021overview}, wireless localization \cite{win2018theoretical,laoudias2018survey}, as well as WiFi and mobile sensing \cite{yassin2016recent,wu2017device,tan2018exploiting}.
These studies present a comprehensive collection of methods and applications relevant to the field.

The growing interest in combining communication and perception has led to key developments in joint sensing-communication (JSC) technology. 
Sturm \textit{et al.}  \cite{sturm2009ofdm} pioneered a method using Orthogonal Frequency Division Multiplexing (OFDM) symbols for JSC signal processing, enhancing radar signal processing while supporting both radar ranging and communication.
Moghaddasi \textit{et al.} \cite{moghaddasi2016multifunctional} introduced a reconfigurable receiver that combines radar sensing and radio communication on a time-division basis, although this method limits simultaneous radar and communication functions.
Zhang \textit{et al.} \cite{zhang2018multibeam} proposed a JSC system for mobile communication using multibeam-forming in a Time-Division-Duplex (TDD) setup.
Significant efforts have also been made in developing the Cooperative Sensing Unmanned Aerial Vehicle Network (CSUN), with studies like Causa \textit{et al.} \cite{causa2018multi} focusing on cooperative navigation and Kanellakis \textit{et al.} \cite{kanellakis2018cooperative} on large infrastructure sensing. 
Hildmann \textit{et al.}\cite{hildmann2019using} explored a CSUN using separate devices for radar and communication. 
However, these approaches, focusing on CSUN, often overlook the integrated potential of JSC techniques, missing out on the opportunity for full spectrum reuse between radar and communication, a limitation this paper refers to as conventional CSUN.

The integration of perception/sensing and communication technologies, particularly through radar systems, has been a subject of extensive research, leading to significant advancements and insights in the field.
Wang \textit{et al.} \cite{wang2021overview} reviewed radar target detection methods, shedding light on modern radar system complexities. 
Win \textit{et al.} \cite{win2018theoretical} delved into the fundamentals and challenges of Time-Of-Arrival (TOA) wireless localization, analyzing practical estimation algorithms. 
Paul \textit{et al.} \cite{paul2016survey} discussed the applications, topologies, integration levels, and current advances in joint radar-communication systems, offering insights into future directions. 
Liu \textit{et al.} \cite{liu2020jointradar} reviewed radar-communication coexistence and dual-functional system research, while Zhang \textit{et al.} \cite{zhang2021enabling} focused on the coexistence strategies of radar and communication systems, covering signal models, waveform design, and processing techniques. 
Together, these studies advance the understanding of sensing and communication technology integration, setting the stage for future breakthroughs.

The integration of radar sensing and communication has been extensively explored, yet the integration of visual capture devices into this realm has not been as thoroughly investigated. 
Leveraging insights from our prior research in smart healthcare \cite{ye2024deep}, which explored the synergy between visual and sensory data, this study breaks new ground. 
To our knowledge, the specific use of image recognition to boost communication efficiency remains largely untapped.
This paper adopts camera data to refine computation scheduling and resource distribution in the \systemname{}, inspired by our previous methodologies. 
By merging visual with mmWave data, we aim to not only broaden the spectrum of sensing information, providing deeper environmental understanding, but also to enhance the decision-making process within the system. 
This innovation proves particularly valuable in the dynamic and often unpredictable \systemname{} environment, improving both computation offloading and resource allocation efficiency.

\section{System Model}
\subsection{Network Settings}
\begin{figure*}[!t]
    \centering
    \includegraphics[width=7in]{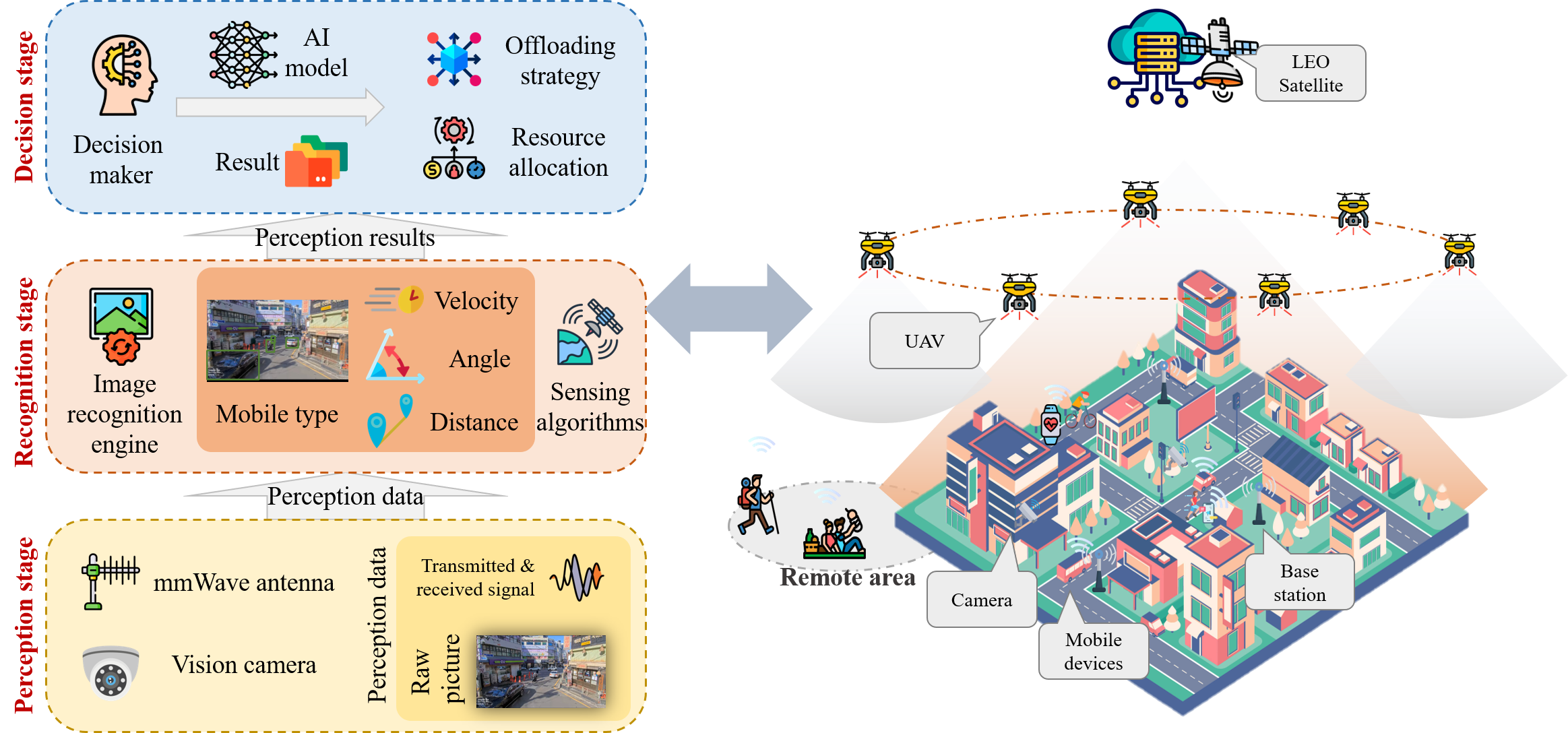}
    \caption{Architecture of \systemname{} and the entire process of deep reinforcement learning and perception-aided approach for \systemname{}.}
    \label{fig:1}
\end{figure*}

This section outlines the network formulation and problem statement.
As shown in Figure \ref{fig:1}, the \systemname{} is envisioned as a comprehensive platform that integrates all principal communication networks to provide seamless connectivity. 
This paper proposes deploying \systemname{} as a solution to service requests in areas inaccessible to conventional cellular networks. 
It acknowledges that, even in remote locations, there can be periods of high user density, such as during special events or in temporarily populated areas, which require robust service management.
The proposed \systemname{} system is structured around three primary layers: ground, air, and space, each designed to effectively handle service demands from mobile devices, regardless of their location's network infrastructure.
The space component is represented by multiple \leoname{} satellites, tasked with delivering cloud computing services for the targeted area.
The aerial segment incorporates $M$ UAVs that serve as mobile edge nodes, offsetting the lack of ground BS coverage. 
These UAVs are deployed to provide edge computing and caching services to terrestrial mobile devices and are outfitted with solar panels for prolonged operation without the necessity for frequent recharging. 
Although the path followed by UAVs significantly impacts the efficiency of task offloading, our research does not delve into optimizing UAV flight paths. 
We proceed with an established trajectory for the UAVs, ensuring they adequately cover the designated area throughout this paper.
Additionally, the UAVs are equipped with mmWave and vision sensors, as detailed in Section III.D. 

We introduce a configuration of potential surrounding BSs labeled as ${\mathcal N}=\{1, 2, \ldots, n, \ldots, N\}$, and define ${\mathcal K}=\{1, 2, \ldots, k, \ldots, K\}$ as the index set for $K$ mobile devices located within the targeted areas. 
These devices navigate using a random walk model, moving at a velocity of $v_g$.
Their direction of movement periodically changes at a random angle $\sin_g$, yet the velocity remains constant.
The group of UAVs is marked as ${\mathcal M}=\{1, 2, \ldots, m, \ldots, M\}$.
Given the finite battery capacity of mobile devices, they can offload computing tasks not just to the UAVs but also to the \leoname{} satellite, facilitating computational tasks.
A list of mathematical symbols and functions most frequently used in this paper is available in Table \ref{table:1}. 

\subsection{Task Model}
In the \systemname{} system, we utilize a discrete time-slotted framework. 
The series of time slots is denoted as ${\mathcal T}=\{1, 2, \ldots, t, \ldots, T\}$, with the duration of each slot is $\tau$. 
The computation tasks generated by each device $k\in{\mathcal K}$ during time slot $t$ are represented by $D_k(t)$. 
The average task arrival rate for device $k$ in time slot $t$ is denoted as $\lambda_k={\mathbb E}\{D_k(t)\}$.
Each device, depending on the application it runs, requires $\gamma_k$ CPU cycles to process one bit of data.
The approach to computation offloading significantly hinges on the computing task model selected. 
The two main models, binary and partial offloading, cater to different task complexities. 
Binary offloading treats tasks as whole units for either local or server processing, suitable for simple or integrated tasks.
Conversely, partial offloading splits tasks into independent parts for concurrent processing, fitting complex applications in mobile devices in remote areas with sparse ground BS coverage. 
This study specifically employs the partial offloading model at the UAV layer to facilitate intermediary connectivity.
\begin{table}[!t]
\centering
\caption{List of Notations}
\begin{tabular}{|c|p{2.55cm}|c|p{2.55cm}|}
\hline
${\mathcal M}$ & set of UAVs & ${\mathcal N}$ & set of BSs\\
\hline
${\mathcal K}$ & set of ground devices   & ${\mathcal T}$ & set of time slots\\
\hline
$\tau$ &\multicolumn{3}{p{6cm}|}{duration of each time slot}\\
\hline
$\triangle$ &\multicolumn{3}{p{6cm}|}{duration of Phase 1}\\
\hline
\multirow{2}{*}{$L_{k,m}^{ud}(t)$} &\multicolumn{3}{p{6cm}|}{transmission latency from ground device $k$ to its corresponding UAV $m$ in time slot $t$}\\
\hline
\multirow{2}{*}{${\mathcal C}_m(t)$}& \multicolumn{3}{p{6cm}|}{subset of ground devices within the coverage area of UAV $m$ in time slot $t$}\\
\hline
\multirow{2}{*}{${\mathcal BS}_m(t)$}& \multicolumn{3}{p{6cm}|}{subset of BSs within the coverage area of UAV $m$ in time slot $t$}\\
\hline
{ $\mbox{Loc}_k(t)$ }&\multicolumn{3}{p{6cm}|}{  Location of ground device $k$ in time slot $t$}\\
\hline
\multirow{2}{*}{${\boldsymbol x}(t) $}&\multicolumn{3}{p{6cm}|}{ task hosting decision associated all UAVs in time slot $t$, ${\boldsymbol x }(t)=\{{\boldsymbol x}_m\}_{m\in{\mathcal M}}$}\\
\hline
\multirow{2}{*}{${\boldsymbol Q}(t)$} &\multicolumn{3}{p{6cm}|}{computation offloading decision in time slot $t$, ${\boldsymbol Q}(t)=\{[Q_m^{u,loc}(t), Q_m^{u,b}(t), Q_m^{u,s}(t)]\}_{m\in{\mathcal M}}$ }\\
\hline
\multirow{2}{*}{${\boldsymbol f}^u(t)$ }& \multicolumn{3}{p{6cm}|}{computation resource allocation of UAVs in time slot $t$, ${\boldsymbol f}^u(t)=(f_1^u(t), \ldots, f_M^u(t))$}\\
\hline
\multirow{2}{*}{${\boldsymbol f}^{bs}(t)$} & \multicolumn{3}{p{6cm}|}{computation resource allocation of BSs in time slot $t$, ${\boldsymbol f}^{bs}(t)=(f_{m,n}^{bs}(t), m\in{\mathcal M}, n\in{\mathcal N})$}  \\
  \hline
\multirow{2}{*}{${\boldsymbol H}(t)$} &\multicolumn{3}{p{6cm}|}{state of \systemname{} in time slot $t$, ${\boldsymbol H}=\{H_m^{u}(t), H_{m,n}^{u,b}(t)\}_{m\in{\mathcal M}, n\in{\mathcal N}}$}\\
\hline
\multirow{2}{*}{$J_m^{u,\bigstar}(t)$}&\multicolumn{3}{p{6cm}|}{quantities of tasks processed in time slot $t$ for $Q_m^{u,\bigstar}, \forall \bigstar\in\{loc, b, s\}$ }\\
\hline
\multirow{2}{*}{$C_m^{u,\bigstar}(t)$} &\multicolumn{3}{p{6cm}|}{cost of computation models \textcircled{1}, \textcircled{2}, and \textcircled{3}, with $\textcircled{1}\leftrightarrow loc, \textcircled{2}\leftrightarrow b, \textcircled{3}\leftrightarrow s, \forall \bigstar\in\{loc, b, s\}$}\\
\hline
\multirow{3}{*}{$E_m^{\bigstar,cmp}(t)$} &\multicolumn{3}{p{6cm}|}{energy consumed at $\bigstar, \forall \bigstar\in\{loc, b, s\}$ in time slot $t$, where the computation offloading is assigned by UAV $m$}\\
\hline
\multirow{2}{*}{$E_m^{u,\bigstar,tran}(t)$ }  &\multicolumn{3}{p{6cm}|}{ energy consumed for UAV $m$ to offload tasks via UAV-$\bigstar$ link, with $\bigstar\in\{b, s\}$ }\\
\hline
$C_k^{d,s,dir}(t)$ &\multicolumn{3}{p{6cm}|}{cost of direct offloading to the LEO satellite}\\
\hline
\end{tabular}
\label{table:1}
\end{table}
\subsection{Communication Model}
Given the volume of tasks it receives and its limited computational capacity, the UAV determines whether to offload tasks to \leoname{} for cloud server processing or to a nearby BS. 
As it selects the offloading destination, the UAV simultaneously determines how to segment the tasks, aiming for an optimal distribution of the workload. 
The process involves two transmission links, air-to-ground and air-to-satellite.
We proceed to detail these communication channel models, emphasizing the UAV layer's pivotal role as the central control layer within \systemname{}.

\subsubsection {Communication Rate Between Air and Ground}
Building on the work \cite{chen2021energy}, we assume that the communication link between UAV $m$ and BS $n$ operates within the C-band frequency spectrum. 
The channel coefficient $h_{m,n}^{ub}(t)$, representing this connection at any given time slot $t$, is defined as
\begin{align}\label{eq:1}
h_{m,n}^{ub}(t)=\epsilon_{m,n}^{ub}(t)\beta \sqrt{1/PL(d_{m,n}^{ub}(t))},
\end{align}
$\epsilon_{m,n}^{ub}(t)$ is the small-scale fading between UAV $m$ and BS $n$ at time slot $t$, which adheres to a Rayleigh distribution, specifically $\epsilon_{m,n}^{ub}(t)\sim \mathcal{CN}(0,1)$.
The term $\beta$ represents shadow fading, and $d_{m,n}^{ub}(t)$ indicates the distance between UAV $m$ and BS $n$ at time slot $t$.
The function $PL(d_{m,n}^{ub}(t))$ quantifies the large-scale path loss. 
Although the distance between the UAV and the BS varies over time, for simplicity, we assume that the UAV's position remains constant within a single time slot, using its location at the beginning of the slot.

Then, the transmission rates in bits/second for links from UAV $m$ to BS $n$ and from BS $n$ to UAV $m$ at time slot $t$ are specified as follows:
\begin{align}
R_{m,n}^{ub}(t)&=B_{m,n}^{ub}\log_2\left(1+\frac{P_{m,n}^{ub}|h_{m,n}^{ub}(t)|^2}{N_0B_{m,n}^{ub}}\right), \label{eq:2}\\
R_{n,m}^{ub}(t)&=B_{m,n}^{ub}\log_2\left(1+\frac{P_{n,m}^{bu}|h_{m,n}^{ub}(t)|^2}{N_0B_{m,n}^{ub}}\right),\label{eq:3}
\end{align}
where $B_{m,n}^{ub}$ represents the bandwidth allocated to the link, $P_{m,n}^{ub}$ denotes the transmission power from UAV $m$ to BS $n$, and $P_{n,m}^{ub}$ indicates the transmission power from the BS to the UAV. Furthermore, $N_0$ refers to the Gaussian noise power spectral density within the terrestrial network.

Define ${\mathcal C}_m(t)=\{k\in{\mathcal K} ~|~\mbox{Loc}_k(t)\in {\mathbb R}_m(t)\}, m\in{\mathcal M}$, be the subset of devices in the coverage area ${\mathbb R}_m(t)$ of UAV $m$ at time slot $t$, such that $\cap_{m\in{\mathcal M}} {\mathcal C}_m(t)=\emptyset$.
$\mbox{Loc}_k(t)$ represents the location of ground device $k$ at time slot $t$.
It is noted that this paper does not emphasize meeting the condition $\cup_{m\in{\mathcal M}}{\mathcal C}_m(t)={\mathcal K}$.
In other words, user mobility results in a fluctuating presence within the area, leading to temporal variations in the number of users serviced by all UAVs.
At time slot $t$, the transmission rate from UAV $m$ to its covered ground device $k$ is indicated as $R_{m,k}^{ud}(t)$, while the transmission rate from the $k$ to UAV $m$ is represented as $R_{k,m}^{ud}(t)$. 
These rates can be derived using Eqs. \eqref{eq:2} and \eqref{eq:3}. It's important to note that the distance between UAV $m$ and the ground device $k$ it serves during the time slot is calculated using mmWave perception technology, denoted as $d_{m,k}^{ud}(t)$. 
The specifics of this calculation process are detailed in Section III.D.

\subsubsection {Communication Rate Between Air and Satellite}
The UAV to satellite link primarily features a clear Line of Sight (LoS) path, complemented by minor Non-Line of Sight (NLoS) scattered components. 
This configuration leads us to model the UAV-Satellite channel as Rician \cite{di2019ultra, abdi2003new}.
Consequently, the channel gain, integrating both LoS and NLoS elements, is shown as 
\begin{align}\label{eq:4}
h_{0,m}^{us}(t)=\sqrt{\frac{F}{1+F}}h_{0,m}^{us,LoS}(t)+\sqrt{\frac{(d_{0,m}^{us}(t))^{-\alpha_1}}{1+F}}h_{0,m}^{us,NLoS}(t),
\end{align}
where $h_{0,m}^{us, LoS}(t)$ denotes the channel gain for LoS scenario, expressed as 
\begin{align}\label{eq:5}
h_{0,m}^{us,LoS}(t)=\sqrt{(d_{0,m}^{us}(t))^{-\alpha_2}}e^{-j\frac{2\pi}{\lambda}d_{0,m}^{us}(t)}.
\end{align}

Here, $d_{0,m}^{us}(t)$ represents the distance between UAV $m$ and the satellite during time slot $t$, with $\alpha_2$ indicating the pathloss exponent for LoS communication and $\lambda$ being the wavelength in the Ka-band.  Given that the satellite's orbit is predetermined, its longitudinal, latitudinal, and altitudinal coordinates are known for each time slot, and it is assumed that these coordinates remain stable throughout the duration. 
$\alpha_1$ refers to the pathloss exponent for NLoS transmission, while $h_{0,m}^{us,NLoS}(t)$ denotes the associated small-scale channel gain, adhering to the distribution $h_{0,m}^{us,NLoS}(t)\sim\mathcal{CN}(0,1)$.
Additionally, $F$ is the Rician factor. 

Subsequently, the transmission rates achievable for communications from UAV $m$ to the satellite, denoted by $R_{m,0}^{us}(t)$, and for communications from the satellite to UAV $m$, represented by $R_{0,m}^{us}(t)$, are determined as follows:
\begin{align}
R_{m,0}^{us}(t)&=B_{0,m}^{us}\log_2\left(1+\frac{P_{m,0}^{us}G_0|h_{0,m}^{us}(t)|^2}   {N_0B_{0,m}^{us}}\right),\label{eq:6}\\
R_{0,m}^{us}(t)&=B_{0,m}^{us}\log_2\left(1+\frac{P_{0,m}^{us}G_0|h_{0,m}^{us}(t)|^2}{N_0B_{0,m}^{us}}  \right), \label{eq:7}
\end{align}
where $B_{0,m}^{us}$ refers  to the bandwidth available in the Ka-band, $G_0$ is the fixed  antenna gain, while $P_{m,0}^{us}$ and $P_{0,m}^{us}$ specify the transmission powers from UAV $m$ to the satellite and from the satellite back to UAV $m$, respectively.

Furthermore, the transmission rates from any ground device $k$ to the satellite, denoted as $R_{k,0}^{ds}(t)$, along with the rates from the satellite back to ground device $k$, denoted as $R_{0,k}^{ds}(t)$, are calculated based on Eqs. \eqref{eq:6} and \eqref{eq:7}. 
Following this, the next section delves into the development of a dynamic state perception model, which leverages both visual and mmWave data to enhance computation offloading decision-making process.

\subsection{Dynamic State Perception Model}
We propose a dynamic state perception model that integrates cameras with millimeter-wave radar, effectively compensating for the radar's limitations in capturing target shape and texture information.  
In this setting, UAVs are equipped with an array of cameras and mmWave sensor systems. 
This equipment significantly enhances their ability to detect and identify the movements and behaviors of various entities, including pedestrians, cyclists, and vehicles, especially in remote areas where conventional surveillance and data collection techniques are constrained or non-existent.
These vision sensors are tasked with capturing real-time imagery, which is then processed by an {\tt YOLOv7} \cite{wang2023designing,wang2023yolov7} designed to complement the radar's detection capabilities.
As illustrated in Fig. \ref{fig:1}, once subjects are detected by mmWave radar, visual capture devices are deployed to further classify and analyze the categories and behaviors of the detected objects, including their state, movement speed, and direction.
This integration of visual recognition and radar detection significantly refines the decision-making process. 
It ensures that the combined efforts in task offloading, resource allocation, and association control are optimized, accommodating the complex environment characterized by uncertain entity behaviors and a fluctuating user base.
The overall goal of this Dynamic Status Perception Model is to leverage cutting-edge AI methodologies for the nuanced interpretation of environment dynamics, thereby refining network operations and resource distribution within the \systemname{} infrastructure.

mmWave radar utilizes high-frequency waveforms to achieve precise location tracking and accurate distance estimation. 
At the heart of this technology is the transmission of a specific waveform type, known as a linear Frequency-Modulated Continuous Wave (FMCW). 
This waveform is distinguished by a frequency that increases linearly over time, offering a robust method for measuring both the distance and relative speed of objects within its detection range.
Inspired by \cite{gao2021mimo}, the transmitted signal can be expressed as
\begin{align}\label{eq:loc1}
    s_T(t)=A_T\cos[2\pi(f_0t+St^2/2)+t_0],
\end{align}
where $A_T$ is the amplitude of the transmitted signal, $f_0$ is the center frequency, $S$ is the modulation frequency and $S=B/T$.
$B$ is the transmitted signal sweep bandwidth, $T$ is signal frequency raising cycle. $t_0$ is the initial phase.

To elucidate the mechanism by which mmWave radar systems ascertain the distance to a device, consider a device $k\in{\mathcal K}$ positioned at a distance $d_{m,k}^{ud}$ from radar $m$ (i.e., UAV $m$). 
As electromagnetic waves travel through the air at the constant speed of light, $c$, the signal that reflects back to the radar's receiving antenna after encountering the device can be mathematically modeled. 
This returned signal retains the characteristics of the initial transmission but exhibits a time delay that corresponds to the distance traveled, detailed as follows:
\begin{align}\label{eq:loc2}
    s_R(t) = A_R\cos\{2\pi[f_0(t-\tau_0)+S(t-\tau_0)^2/2]+t_0\},
\end{align}
where $A_R$ is the amplitude of the received signal, $\tau_0 = \frac{2d_{m,k}^{ud}}{c}$ is the fixed delay of the received signal relative to the transmitted signal.
Upon reception, the radar system mixes the transmitted signal with this received signal to derive an intermediate frequency (IF) signal.
This process effectively isolates the delay-induced frequency shift, which is directly tied to the distance of interest. 
The IF signal is represented as:
\begin{align}\label{eq:loc3}
    s_{IF}=\frac{1}{2}A_TA_R\cos(2\pi f_0\tau_0+2\pi S\tau_0 t-\pi S \tau_0^2),
\end{align}
and the frequency of the intermediate frequency signal is
\begin{align}\label{eq:loc4}
    f_{m,k}=S\cdot \tau_0 = \frac{S2d_{m,k}^{ud}}{c} = \frac{2Bd_{m,k}^{ud}}{cT}.
\end{align}
We can estimate the distance based on the frequency information of the echo signal
\begin{align}\label{eq:loc5}
    d_{m,k}^{ud}=\frac{cTf_{m,k}}{2B}.
\end{align}

In a scenario where device $k$ continuously emits signals from a fixed distance relative to the radar, the returned echo signals exhibit phase variations. 
These variations are attributed to the Doppler effect, which arises from the motion of the device. 
By meticulously analyzing the phase differences between consecutive echo signals, the velocity of the device can be precisely calculated.
The velocity $v_{m,k}$ is derived by
\begin{align}\label{eq:loc6}
    v_{m,k}=\frac{\lambda \omega_{m,k}}{4\pi T_s},
\end{align}
where $\omega_{m,k}$ is the angular frequency of echo signal, $\lambda$ is echo wavelength and $T_s$ the interval between two adjacent echo signals.

The angular displacement $\theta_{m,k}$ is related to the distance differential between device $k$ and the individual antennas of UAV $m$, which induces a phase difference that can be translated into an angle through the equation:
\begin{align}\label{eq:loc7}
    \theta_{m,k} = \sin^{-1}\left ( \frac{\lambda \omega_{m,k}}{2\pi d_{m,k}^{ud}}\right ).
\end{align}

\section{Models for Computation Offloading and Associated Cost Analysis}
In this study, we designate the UAV layer as the control layer, primarily due to its pivotal role in minimizing latency within the \systemname{}.
We structure each time slot to encompass two phases, reflecting the computational scheduling process. 
This division is essential for smoothly coordinating the transfer and processing of data, making sure that UAVs efficiently direct the exchange of information and computing tasks between ground devices and either the satellite or BSs.
This approach not only streamlines operations but also significantly enhances the overall efficiency and responsiveness of the network, particularly in remote areas where conventional communication infrastructures might be lacking.
In particular, Phase 1 focuses on deciding if the tasks $D_k(t)$ from ground device $k \in{\mathcal C}_m(t)$ can be hosted by its associated UAV $m$ at time slot $t$.
During this initial phase, UAVs leverage their onboard vision and mmWave radar sensors to accurately assess the condition of ground devices, including distance, behaviors, and movement speed and direction, etc.  
The duration of Phase 1 is denoted by $\triangle$.
Following this, Phase 2 focuses on refining the offloading decisions by the UAVs, drawing from the results of task hosting established in Phase 1.

\subsection{Computation Offloading Model}
During Phase 1, an indicator function $x_{k,m}(t)$, as defined in \eqref{eq:8}, is used for each ground device $k$ within ${\mathcal C}_m(t)$. 
This function is set to $1$ if UAV $m$ can host the task $D_k(t)$ from ground device $k$, and it is set to $0$ if not.  
\begin{align}\label{eq:8}
x_{k,m}(t)=\begin{cases} 1, \text{~if~} L_{k,m}^{ud}(t)\leq\triangle \&\& v_k^d(t)\leq \bar{v}^d,\\
0, \text{~otherwise},
\end{cases}
\end{align}
where $L_{k,m}^{ud}(t)=\frac{D_k(t)}{R_{k,m}^{ud}(t)}$ denotes the transmission latency from ground device $k$ to its corresponding UAV $m$ during time slot $t$, while $\bar{v}^d$ represents the maximum speed threshold for the movement of ground devices.

When $x_{k,m}(t)=1$, this indicates that the task $D_k(t)$ from ground device $k\in{\mathcal C}_m(t)$ is hosted by UAV $m$ during time slot $t$. As a result, within this offloading framework, the total volume of tasks hosted by UAV $m$ is 
\begin{align}\label{eq:10}
D_m^{u}(t)=\sum\nolimits_{k\in{\mathcal C}_m(t)}x_{k,m}(t) D_k(t).
\end{align}
Based on \eqref{eq:10},  we introduce the following three offloading models. 
It is assumed that each UAV possesses a sufficient task buffer.
For a given UAV $m\in{\mathcal M}$ during time slot $t$, we identify three key parameters: $Q_m^{u,loc}(t)$, which represents the tasks processed locally by UAV $m$; $Q_m^{u,b}(t)$, the tasks offloaded to neighboring BSs; and $Q_m^{u,s}(t)$, the tasks offloaded to the LEO satellite. 
The quantities of tasks processed in time slot $t$ for $Q_m^{u,loc}(t)$, $Q_m^{u,b}(t)$, and $Q_m^{u,s}(t)$ are $J_m^{u,loc}(t)$, $J_m^{u,b}(t)$, and $J_m^{u,s}(t)$, respectively. 
These offloading models are defined from the UAVs' viewpoint, indicating that tasks from ground devices are hosted by UAVs.
 
{\bf \textcircled{1} $\xrightarrow[x_{k,m}(t)=1]{\text{UAV}}$ UAV.} 
Given the finite computational resources of UAVs, it's crucial to manage the queue backlog when tasks are hosted to UAV. 
We denote the computation queue of UAV $m$ during time slot $t$ as $H_m^u(t)$. 
To enhance system performance, UAVs utilize a frequency scaling technique\cite{xu2018uav}, allowing for the dynamic adjustment of the CPU cycle frequency. 
The CPU cycle frequency for UAV $m$ at time slot $t$ is indicated by $f_m^u(t)$, subject to a maximum limit of $f_{m,\max}^u$. 
This approach ensures that UAVs can adapt their processing capabilities within their operational constraints to efficiently handle the incoming tasks.
As a result, the queuing dynamics of $H_m^u(t)$ and the volume of tasks processed, denoted by $J_m^{u,loc}(t)$, are modeled as follows:
\begin{align}
H_m^u(t+1)=&\max\left\{H_m^u(t)-J_m^{u,loc}(t), 0\right\} \notag\\ 
&-Q_m^{u,b}(t)-Q_m^{u,s}(t)+D_m^u(t),\label{eq:11}\\
J_m^{u,loc}(t)=&\min\left\{H_m^{u}(t), {f_m^u(t)(\tau-\triangle)}/ {\gamma_m^u}\right\},\label{eq:12}
\end{align}
where $\gamma_m^u$ represents the quantity of CPU cycles required to process one bit of data.

{\bf \textcircled{2} $\xrightarrow[x_{k,m}(t)=1]{\text{UAV}}$ BS.}
Let $\mathcal{BS}_m(t)=\{n\in{\mathcal N} | \mbox{Loc}_n\in{\mathbb R}_m(t)\}$, denote the subset of BSs within the coverage area of ${\mathbb R}_m(t)$ of UAV $m$ during time slot $t$. 
Let $y_{m,n}(t)$ represent the link establishment between UAV $m$ and a BS $n\in\mathcal{BS}_m(t)$, where $y_{m,n}(t)\in\{0, 1\}$.
A value of $y_{m,n}(t)=1$ signifies that UAV $m$ has selected BS $n\in\mathcal{BS}_m(t)$ for task offloading during time slot $t$; otherwise, $y_{m,n}(t)=0$ indicates no offloading to BS $n$. 
Consequently, the volume of tasks processed at the nearby BS is determined by
\begin{align}\label{eq:13}
Q_m^{u,b}(t)=\sum_{n\in\mathcal{BS}_m(t)}y_{m,n}(t)\min\left\{ \tilde{H}_m^u(t), (\tau-\triangle)R_{m,n}^{ub}(t)\right\},
\end{align}
where $\tilde{H}_m^u(t)=\max\left\{H_m^u(t)-Q_m^{u,loc}(t), 0\right\}-Q_m^{u,b}(t)-Q_m^{u,s}(t)+D_m^u(t)$. 

We assume that each BS is equipped with task buffers of sufficient capacity to accommodate tasks from $M$ UAVs.
Here, denote $H_{m,n}^{u,b}(t)$ as the computation queue at BS $n\in\mathcal{BS}_m(t)$, for tasks sent by UAV $m$ within time slot $t$.
The tasks executed from $H_{m,n}^{u,b}(t)$ in this period are represented by $J_{m,n}^{u,b}(t)$. 
We note the computational power assigned by BS $n\in\mathcal{BS}_m(t)$ to UAV $m$ as $f_{m,n}^{u,b}(t)$, which is constrained by $f_{n,\max}^b(t)$. Therefore, we explore the dynamics of the queue $H_{m,n}^{u,b}(t)$ and the processed tasks $J_{m,n}^{u,b}(t)$ in our forthcoming analysis, detailed as follows:
\begin{align}
H_{m,n}^{u,b}(t+1)=&\max\left\{H_{m,n}^{u,b}(t)-J_{m,n}^{u,b}(t), 0\right\} \notag \\
&+y_{m,n}(t)Q_m^{u,b}(t), \label{eq:14}\\
J_{m,n}^{u,b}(t)=&\min\left\{H_{m,n}^{u,b}(t), {(\tau-\triangle)f_{m,n}^{bs}(t)}/{\gamma_m^u}  \right\}.\label{eq:15}
\end{align}

{\bf \textcircled{3} $\xrightarrow[x_{k,m}(t)=1]{\text{UAV}}$  Satellite.}
Similar to task processing at the BSs, the quantity of tasks successfully completed is determined as follows:
\begin{align}\label{eq:16}
J_{m}^{u,s}(t)=\min\left\{H_m^u(t), (\tau-\triangle)R_{m,0}^{us}(t)\right\}.
\end{align}
Commonly, the cloud server on the \leoname{} satellite is equipped with a multi-core CPU. 
We assume that tasks directed to the \leoname{} satellite are processed instantly, eliminating any queue backlogs and queueing delays.

{\bf \textcircled{4} Ground Device $\xrightarrow[x_{k,m}(t)=0]{\text{direct}}$ Satellite:} 
When $x_{k,m}(t)=0$, it signifies that ground device $k\in{\mathcal C}_m(t)$ is scheduled to offload its tasks to the satellite during time slot $t$. 
Consequently, in this offloading scenario, the volume of tasks directed to the satellite is
\begin{align}\label{eq:9}
Q_0^{s,dir}(t)=\sum\nolimits_{m\in{\mathcal M}}\sum\nolimits_{k\in{\mathcal C}_m(t)}[1-x_{k,m}(t)]D_k(t).
\end{align}

\subsection{Cost Model}
The primary goals of computation offloading and resource scheduling in the \systemname{} are centered on minimizing the network's operational cost.
To this end, we have developed corresponding cost models for the four  computation offloading models, considering both the energy consumption and server usage costs involved.
In the following, we will initially explore the three cost models associated with the computation offloading models, viewed from the UAV's perspective. 
Specifically, when $x_{k,m}(t)=1$, it indicates that the UAV undertakes the management and hosting of tasks originating from devices.

\subsubsection{Cost of Computation Offloading Model \textcircled{1}}
During time slot $t$, the energy consumed by UAV $m$ for local computation, denoted as $E_m^{u,cmp}(t)$, is
\begin{align}\label{eq:21}
E_m^{u,cmp}(t)=\kappa[f_m^u(t)]^3\frac{\gamma_m^u H_m^{u}(t)}{f_m^u(t)} ,
\end{align}
where $\kappa$ represents the effective switching capacitance of the CPU, a factor determined by the CPU's hardware architecture. 
Since local computing bypasses the need for BSs or LEO satellites, there is no server usage cost involved.
Consequently, the cost associated with local computation for UAV $m$ is 
\begin{align}\label{eq:22}
C_m^{u,loc}(t)=E_m^{u,cmp}(t).
\end{align}

\subsubsection{Cost of Computation Offloading Model \textcircled{2}}
In time slot $t$, the energy required for UAV $m$ to offload tasks to nearby BSs is derived by  
\begin{align}\label{eq:23}
E_m^{u,b,tran}(t)=\sum_{n\in\mathcal{BS}_m(t)} y_{m,n}(t) P_{m,n}^{ub}(t)\frac{Q_m^{u,b}(t)}{R_{m,n}^{ub}(t)}.
\end{align}

The energy consumed at BSs associated with UAV $m$ in time slot is 
\begin{align}\label{eq:24}
E_m^{u,b,cmp}(t)=&\sum_{n\in\mathcal{BS}_m(t)} \kappa y_{m,n}(t) |f_{m,n}^{bs}(t)|^3
\frac{\gamma_m^u\cdot H_{m,n}^{u,b}(t)}{f_{m,n}^{bs}(t)}.
\end{align}
Thus, the cost of computation offloading model \textcircled{2} is 
\begin{align}\label{eq:25}
C_m^{u,b}(t)=E_m^{u,b,tran}+E_m^{u,b,cmp}(t).
\end{align}

\subsubsection{Cost of Computation Offloading Model \textcircled{3}}
Similarly, the energy consumed for UAV $m$ to offload tasks via UAV-satellite link is 
\begin{align}\label{eq:26}
E_m^{u,s,tran}(t)=P_{m,0}^{us}(t)\frac{Q_{m}^{u,s}(t)}{R_{m,0}^{us}(t)}.
\end{align}

Drawing inspiration from \cite{liu2021joint}, the computation resource usage of \leoname{} satellite by UAV $m$ is quantified by
\begin{align}\label{eq:27}
E_m^{u,s,cmp}(t)=Q_m^{u,s}(t)\gamma_m^u.
\end{align}
Thus, the cost of computation offloading model \textcircled{3} is 
\begin{align}\label{eq:28}
C_m^{u,s}(t)=E_m^{u,s,tran}(t)+E_m^{u,s,cmp}(t).
\end{align}

\subsubsection{Cost of Computation Offloading Model \textcircled{4}}
Define $P_{k,0}^{ds}(t)$ as the transmission power for ground device $k$. The energy consumed to offload tasks from device $k$ to the satellite via the device-to-satellite link is given by $E_k^{d,s,tran}(t) = P_{k,0}^{ds}{D_k(t)}/{R_{k,0}^{ds}(t)}$.
Likewise, the computation resource usage of the \systemname{} satellite by ground device $k$, denoted as ${{E}}_{k}^{d,s,cmp}(t)$, is quantified by ${{E}}_k^{d,s,cmp}(t)=D_k(t)\cdot \gamma_k$.

Therefore, the cost of direct offloading to the \leoname{} satellite is determined as follows: 
\begin{align}\label{eq:20}
C_k^{d,s,dir}(t)=E_k^{d,s,tran}(t)+{{E}}_k^{d,s,cmp}(t).
\end{align}

Additionally, in the Phase 1 of each time slot $t$, the energy consumed by UAV $m$ for collecting tasks via the Device-UAV link is represented as $E_m^{d,u,col}(t)$, which is calculated as
\begin{align}\label{eq:29}
C_m^{d,u,col}(t)=\sum_{k\in{\mathcal C}_m(t)} x_{k,m}(t)P_{k,m}^{d,u}(t)\frac{D_k(t)}{R_{k,m}^{d,u}(t)}.
\end{align}

To conclude, for time slot $t$, considering UAV $m$ and its coverage area ${\mathcal C}_m(t)$, the operational cost attributed to UAV $m$ can be described as
\begin{align}\label{eq:30}
C_m^u(t)=&C_m^{d,u,col}(t)+C_m^{u,loc}(t)+C_m^{u,b}(t)+C_m^{u,s}(t)\\ \notag
&+\sum\nolimits_{k\in{\mathcal C}_m(t)}(1-x_{k,m}(t))C_k^{d,s,dir}(t).
\end{align}

\section{Problem Formulation and Algorithm Design}
This section starts by presenting the formulation of the problem.
It then proceeds to reformulate the problem using Lyapunov optimization for easier resolution.

\subsection{Problem Formulation}
This section is dedicated to reducing the long-term average network operational cost, while also maintaining the stability of the \systemname{}.
We formulate the joint optimization of task hosting ${\boldsymbol x}(t)=\{{\boldsymbol x}_m(t)\}$, where ${\boldsymbol x}_m(t)=(x_{k_1,m}(t), x_{k_2,m}(t), \ldots, x_{k_{|{\mathcal C}_m(t)|},m}(t)), \forall m\in{\mathcal M}$, optimization of computation offloading ${\boldsymbol Q}(t)=\{Q_m^{u,loc}(t), Q_m^{u,b}(t), Q_m^{u,s}(t)\}, \forall m\in{\mathcal M}$, association control (in the perspective of UAVs) ${\boldsymbol y}(t)=\{y_{m,n}(t)\}, n\in{\mathcal N}\cup \{0\}$, where $0$ representing LEO satellite; computing resource allocation of UAVs ${\boldsymbol f}^u(t)=(f_1^u(t), \ldots, f_M^u(t))$, and BS computing resource allocation ${\boldsymbol f}^{bs}(t)=(f_{m,n}^{bs}(t)), m\in{\mathcal M}, n\in{\mathcal N}$. 
The optimization problem is formulated as follows. 
\begin{subequations}\label{eq:31}
\begin{align}
&\min_{{\boldsymbol x}(t), {\boldsymbol y}(t),{\boldsymbol Q}(t), {\boldsymbol f}^u(t), {\boldsymbol f}^{bs}(t)}\lim_{T\rightarrow \infty}\frac{1}{T}\sum_{t\in{\mathcal T}}\sum_{m\in{\mathcal M}}C_m^u(t) \label{eq:31a}\\
\text{s.t.~}& Q_m^{u,loc}(t)+Q_m^{u,b}(t)+Q_m^{u,s}(t)\leq H_m^u(t),\label{eq:31b}\\
&f_m^u(t)\leq f_{m,\max}^u(t), \forall m\in{\mathcal M}, \label{eq:31c}\\
& \sum\nolimits_{m\in{\mathcal M}}f_{m,n}^{bs}(t)\leq f_{n,\max}^{bs}(t), \forall n\in{\mathcal N}, \label{eq:31d}\\
& x_{k,m}(t)\in\{0, 1\}, \forall k\in{\mathcal K}, m\in{\mathcal M},\label{eq:31e}\\
& \sum\nolimits_{m\in{\mathcal M}} x_{k,m}(t)\leq 1, \forall k\in{\mathcal K}, \label{eq:31f}\\
& y_{m,n}(t)\in\{0, 1\}, \forall m\in{\mathcal M}, n\in{\mathcal N}, \label{eq:31g}\\
& \sum\nolimits_{n\in{\mathcal N}} y_{m,n}(t)\leq 1, \forall m\in{\mathcal M},\label{eq:31h}\\
& \bar{H}_m^u(t)< \infty, \bar{H}_{m,n}^{u,b}(t)<\infty, \forall m\in{\mathcal M}, n\in{\mathcal N}, \label{eq:31i}
\end{align}
\end{subequations}
\eqref{eq:31b} indicates that the limit of the task hosting for each UAV. \eqref{eq:31c} specifies the maximum computational capacity of a UAV. \eqref{eq:31d} outlines that the computational capability offloaded to a BS is constrained by its available computing resource. 
\eqref{eq:31e}-\eqref{eq:31h} depict the condition where each ground device can select only one UAV or satellite for task hosting, and every chosen UAV can connect with only one ground BS in any given time slot. 
\eqref{eq:31i} ensures the network stability.

\subsection{Problem Transformation by Lyapunov Optimization}
The challenge presented by problem \eqref{eq:31} stems from its nature as a stochastic optimization problem, complicated by the interdependencies among task hosting, computation offloading, association control, and resource distribution.
To address this, we employ the Lyapunov optimization algorithm, which simplifies the complexity by breaking down the overarching multi-slot stochastic problem into tractable one-slot problems, tackled sequentially.
Assuming the \systemname{}'s state at the current time slot $t$ is represented by ${\boldsymbol H}(t)=\{H_m^u(t), H_{m,n}^{u,b}(t)\}$, we define the Lyapunov function as
\begin{align}\label{eq:32}
{\mathcal L}({\boldsymbol H}(t))=\frac{1}{2}\sum_{m\in{\mathcal M}}\left[H_m^u(t)^2+\sum\nolimits_{n\in{\mathcal N}}H_{m,n}^{u,b}(t)^2\right].
\end{align}
A decrease in ${\mathcal L}({\boldsymbol H}(t))$ signifies a reduction in the task queue backlog, indicating that tasks have been processed locally by UAVs or offloaded during time slot $t$. 
Following this, we introduce the definition of the Lyapunov drift function.
\begin{align}\label{eq:33}
\triangle {\mathcal L}({\boldsymbol H}(t))={\mathbb E}\{{\mathcal L}({\boldsymbol H}(t+1))-{\mathcal L}({\boldsymbol H}(t))~|~{\boldsymbol H}(t)\}.
\end{align}

Here, a small absolute value of $\triangle{\mathcal L}({\boldsymbol H}(t))$ indicates minimal variation in the data queue backlogs across successive time slots.
Achieving minimization of $\triangle{\mathcal L}({\boldsymbol H}(t))$ within each time slot ensures adherence to constraints \eqref{eq:31i}. Furthermore, to reduce operational costs, we derive the drift-plus-penalty by
\begin{align}\label{eq:34}
{\mathcal F}({\boldsymbol H}(t))=\triangle{\mathcal L}({\boldsymbol H}(t))+V\cdot{\mathbb E}\{{\mathcal G}(t)~|~{\boldsymbol H}(t)\},
\end{align}
where  $V\geq 0$ serves as the balancing parameter, and ${\mathcal G}(t) = \sum_{m\in{\mathcal M}} C_m^u(t)$ denotes the operational cost linked to UAV $m$ within the \systemname{} in time slot $t$.
Minimizing ${\mathcal F}({\boldsymbol H}(t))$ directly poses challenges. As a result, Theorem \ref{theorem:1} is utilized to establish its theoretical upper limit.
By focusing on reducing the upper bound of ${\mathcal F}({\boldsymbol H}(t))$ to the greatest extent possible, instead of attempting to directly minimize the function itself, we can approach an optimal operational cost for the network.
\begin{theorem}\label{theorem:1}
Given that $V\geq 0$ and considering the network state in time slot $t$, represented by ${\boldsymbol H}(t)$, then
\begin{align}\label{eq:35}
&{\mathcal F}({\boldsymbol H}(t))\leq \Pi 
+\sum_{m\in{\mathcal M}}{\mathbb E}\{H_m^u(t)[D_m^u(t)-J_m^{u,loc}(t) 
-Q_m^{u,b}(t)\notag \\
&-Q_m^{u,s}(t)]| {\boldsymbol H}(t) \}
+\sum_{m\in{\mathcal M}}\sum_{n\in{\mathcal N}}{\mathbb E}\{H_{m,n}^{u,b}(t)[y_{m,n}(t)Q_m^{u,b}(t) \notag\\
&-J_{m,n}^{u,b}(t)]| {\boldsymbol H}(t)\}
+V{\mathbb E}\{{\mathcal G}({\boldsymbol H}(t))| {\boldsymbol H}(t)\},
\end{align}
where $\Pi$ is a positive constant. 
\end{theorem}

Moreover, by applying Lyapunov optimization, we can ease the stability constraints. 
Ignoring the variations of random variables across distinct time slots, our focus shifts towards optimizing the upper limit of ${\mathcal F}({\boldsymbol H}(t))$ in each time slot.

\subsection{DRL-and-Perception-aided Approach}
Based on Theorem \ref{theorem:1}, we introduce three distinct subproblems aimed at minimizing the right-hand side of \eqref{eq:35}. 
These include {\bf{P1}}, focusing on the joint optimization of computation offloading and local computing resource allocation from the UAVs' viewpoint; {\bf{P2}}, dedicated to optimizing the association control between UAVs and BSs, denoted as $y_{m,n}(t)$; and {\bf{P3}}, concerning the allocation of computing resources at ground BSs. 
We proceed to address these subproblems sequentially.

\subsubsection{Jointly Optimization of Task Offloading and UAV Computing Resource Distribution}
\begin{align}\label{eq:36}
{\bf{P1}:} &\min_{{\boldsymbol Q}(t),{\boldsymbol f}^u(t)} \sum_{m\in{\mathcal M}}H_m^{u}(t)\left[D_m^u(t)-Q_m^{u,b}(t)-Q_m^{u,s}(t) \right.\notag\\ 
&\left.-{f_m^u(t)(\tau-\triangle)}/{\gamma_m^u}\right]+\sum_{m\in{\mathcal M}}V\left[C_m^{d,u,col}(t) \right.\notag \\
&\left.+\sum\nolimits_{k\in{\mathcal C}_m(t)}(1-x_{k,m}(t))C_k^{d,s,dir}(t)  \right]\notag\\ 
&+\sum\nolimits_{m\in{\mathcal M}} V\kappa|f_m^u(t)|^3 {\gamma_m^u H_m^{u}(t)}/{f_m^u(t)} \notag\\
&+\sum_{m\in{\mathcal M}}V \left[P_{m,0}^{u,s}(t) \frac{Q_{m}^{u,s}(t)}{R_{m,0}^{us}(t)} +\kappa|f_{m,0}^{u,s}(t)|^3\frac{\gamma_m^u Q_m^{u,s}(t)}{f_{m,0}^{u,s}(t)}\right] \notag \\
&+\sum_{m\in{\mathcal M}}VP_{m}^{ub}(t)\frac{Q_m^{u,b}(t)}{R_m^{ub}(t)} +\sum_{m\in{\mathcal M}} V\kappa|f_m^{bs}(t)|^3(\tau-\triangle)  \\
\text{s.t.~} &Q_m^{u,loc}(t)+Q_m^{u,b}(t)+Q_m^{u,s}(t)=H_m^u(t), \\
&f_m^u(t)\leq f_{m,\max}^u(t), \forall m\in{\mathcal M},\\
& f_m^u(t)(\tau-\triangle)/\gamma_m^u\leq H_m^u(t) \forall m\in{\mathcal M}.
\end{align}

To tackle {\bf{P1}}, we utilize the DDPG algorithm, a model-free, off-policy actor-critic method embedded within a deep learning framework. 
DDPG is particularly well-suited for continuous action spaces, making it an effective solution for problems characterized by high-dimensional, continuous action domains. This approach ensures precise optimization of task offloading and computing resource allocation among UAVs.
We denote the state space of {\bf{P1}} at time slot $t$ as ${\mathcal S}(t)=\{R^{ub}(t),R^{us}(t),D^u(t),\mbox{CT}(t)\}$, where $\mbox{CT}(t)$ represents the recognition results of ground device types via mmWave radar and visual sensors. 
The action space is denoted as $a(t)=\{\boldsymbol{Q}(t),\boldsymbol{f}^u(t)\}$.
The reward is set as $r(t)=-\text{Eq.}\eqref{eq:36}$.

\subsubsection{Optimization of Association Control $y_{m,n}(t)$}
In this section, we determine the UAVs and the ground BSs association control (i.e., $y_{m,n}(t)$) by minimizing {\bf{P2}}. {\bf{P2}} is formulated as follows. 
\begin{align}
{\bf{P2:}}&\min_{y_{m,n}(t)}\sum_{m\in{\mathcal M}}\sum_{n\in\mathcal{BS}_m} \left\{H_{m,n}^{u,b}(t)[y_{m,n}(t)Q_{m}^{u,b}(t)-J_{m,n}^{u,b}(t) ]\right\} \notag \\ 
&+\sum_{m\in{\mathcal M}}\sum_{n\in\mathcal{BS}_m}V y_{m,n}(t)\left[P_{m,n}^{u,b}\frac{Q_m^{u,b}(t)}{R_{m,n}^{ub}(t)} \right. \notag\\ 
&\left.+\kappa|f_{m,n}^{bs}(t)|^3\frac{\gamma_m^u\cdot H_{m,n}^{u,b}(t)}{f_{m,n}^{bs}}\right] \label{eq:a}\\
\text{s.t.~}&y_{m,n}(t)\in\{0,1\}, \forall m\in{\mathcal M}, n\in{\mathcal N},\\
&\sum\nolimits_{n\in\mathcal{BS}_m(t)} y_{m,n}(t)\leq 1, \forall m\in{\mathcal M}.
\end{align}

{\bf{P2}} is a mixed integer nonlinear problem. When $f_{m,n}^{bs}(t)$ is fixed, the optimal $\{y_{m,n}(t)\}_{m\in{\mathcal M},n\in{\mathcal N}}$ can be derived by Deep Q-Network (DQN).
In our investigation of {\bf{P2}}, we utilized the DQN algorithm, a cutting-edge reinforcement learning technique that integrates deep neural networks with a Q-learning framework.
The DQN algorithm is particularly adept at handling high-dimensional state spaces, which makes it a suitable choice for complex decision-making tasks that are characterized by substantial state representations.

Similarly, we denote the state space of {\bf{P2}} at time slot $t$ as $s(t)=\{D^u(t),\boldsymbol{Q}(t),\boldsymbol{f}^u(t)\}$.
The action space is denoted as $a(t)=\{\boldsymbol{y}(t)\}$.
The reward is set as $r(t)=-\text{Eq.}\eqref{eq:a}$.

\subsubsection{Optimization of BS Computing Resource Allocation}
In this section, we determine the BS computing resource allocation by minimizing {\bf{P3}}. {\bf{P3}} is formulated as follows. 
\begin{align}\label{eq:38}
{\bf{P3}:} &\min_{f_{m,n}^{bs}(t)} \sum_{m\in{\mathcal M}}\sum_{n\in\mathcal{BS}_m(t)} \left[-\frac{H_{m,n}^{u,b}(t)(\tau-\triangle)f_{m,n}^{bs}(t)}{\gamma_m^u}  \right. \notag\\
&\left.+\kappa|f_{m,n}^{bs}(t)|^3
\frac{\gamma_m^u H_{m,n}^{u,b}(t)}{f_{m,n}^{bs}(t)}\right]\\
\text{s.t.~}& \sum\nolimits_{m\in{\mathcal M}} f_{m,n}^{bs}(t)\leq f_{n,\max}^{bs}(t), \forall n,\\
& (\tau-\triangle)f_{m,n}^{bs}(t)/\gamma_m^u\leq H_{m,n}^{u,b}(t), \forall m,n. 
\end{align}

Here, we employ the SGHS algorithm to address problem {\bf{P3}}.
Following \cite{omran2008global}, the adjustment of the bandwidth (BW) is demonstrated by Eq. \eqref{eq:39}, where the preset minimum and maximum search scopes of the SGHS algorithm are $\mbox{BW}_{\min}$ and $\mbox{BW}_{\max}$, respectively. 
Harmony Search is a music-inspired optimization algorithm that mimics the improvisation process of musicians. In our context, the algorithm starts by initializing the parameters, including Harmony Memory Size (HMS), Harmony Memory Considering Rate (HMCR), Pitch Adjusting Rate (PAR), and Number of Iterations (NI). 
In this paper, without losing geneality, we let HMCR and PAR obey normal distribution \cite{geem2001new}, i.e., $\mbox{HMCR}\sim{\mathcal N}(\mu_{\mbox{HMCR}},\sigma_{\mbox{HMCR}}^2)$ and $\mbox{PAR}\sim{\mathcal N}(\mu_{\mbox{PAR}},\sigma_{\mbox{PAR}}^2)$, respectively.
The Harmony Memory (HM) is also initialized with feasible solutions, defined by the minimum and maximum values of the harmonies.
More details of the SGHS is shown in Algorithm \ref{algorithm:1}.
\begin{align}\label{eq:39}
 \mbox{BW}(t)=\left\{\begin{array}{ll}
\mbox{BW}_{\max }-\frac{\mbox{BW}_{\max }-\mbox{BW}_{\min }}{\mbox{NI}}2t &  \text { if } \mbox{iter}<\frac{\mbox{NI}}{2}, \\
\mbox{BW}_{\min } &\text { if } \mbox{iter} \geqslant \frac{\mbox{NI}}{2}, 
\end{array}\right.   
\end{align}
\begin{algorithm}[!t]
\SetAlgoLined
\caption{SGHS Algorithm for BS Computing Resoruce Allocation }\label{algorithm:1}
    {\bf Require:} BS's CPU frequency, datasize for each BS, the queue for BS.\\
    
\KwIn{Initialize the algorithm parameters (HMS, HMCR, PAR, NI, $\mbox{BW}_{\min}$, $\mbox{BW}_{\max}$);
Initialize the HM, minimum value and maximum value of harmony}

\While{$\mbox{iter} \leq \mbox{NI}$}
{Update the $\mbox{BW}$ by Eq. \eqref{eq:39};\\
\eIf{$\mbox{rand} \in (0,1) \leq \mbox{HMCR}$}
{choose a value from HM for i\\
\If{$\mbox{rand} \in (0,1) \leq \mbox{PAR}$}
{adjust the value of i by:\\
$a_{i,\text{new}}=a_{i,\text{old}} + \mbox{rand}\in (0,1) \times \mbox{BW}$
}}
{choose a random variable:\\
$a_i = \min + \mbox{rand}\in (0,1)\times(\max-\min)$}
\If{${\bf{P3}}(\text{new harmony solution}) \leq \text{worst}~({\bf{P3}}(\mbox{HM}))$}
{accept the new harmony and replace the worst in HM with it.}
$\mbox{iter}=\mbox{iter} + 1$
}
best = find the current best solution.\\
\KwOut{The BS computing resource allocation results.}
\end{algorithm}

Algorithm \ref{algorithm:2} encapsulates the comprehensive optimization process for addressing problem \eqref{eq:31}, under the proposed \methodname{}. 
The initial phase involves employing \methodname{}, as described in Algorithm \ref{algorithm:2}, to obtain optimal solutions for task hosting, computation offloading, and managing interactions between UAVs and BSs. 
This step also includes determining the optimal distribution of computing resources among UAVs.
Following this, Algorithm \ref{algorithm:1} is applied to fine-tune the computing resource allocation at the BSs, completing the optimization cycle. 
This systematic approach guarantees the achievement of optimal outcomes for task hosting, computation offloading, and computing resource distribution, encompassing the entire optimization scope proposed by \methodname{}.
\begin{algorithm}[!t]
\SetAlgoLined
\caption{\methodname{} Proposed for Joint Task Hosting, Computation Offloading, Association Control, and Computing Resource Allocation in \systemname{} }\label{algorithm:2}
{\bf Require:} UAV, BS, Cloud server's CPU frequency, datasize for each mobile device, the returned signal for each device $s_R$ (by perception).\\
    
\KwIn{Initialize the DDPG model for jointly optimizing computation offloading and UAVs computing resource allocation, initialize the DQN model for the optimization of the UAVs and ground BSs association control, initialize the harmony search algorithm for optimization BS computing resource allocation;}

\For{time slot $t=1$ to $T$}
{Estimate the distance, velocity and angle based on the radar.\\
    \eIf{User is about to leave the coverage area}
{Leave the device to the satellite}
{UAV collects the task and obtain the amount of tasks by \eqref{eq:10}
}
\For{iteration number $iter=1$ to $ITER$}
{Send state space to the DDPG model to get $\boldsymbol Q(t),\boldsymbol f^u(t)$ for minimize the Problem P1;\\
Send state space to the DQN model to get the $y_{m,n}(t)$ for minimize the Problem P2;\\
Obtain the $f_{m,n}^{bs}(t)$ by Algorithm \ref{algorithm:1};\\
\If{The new solution is no better than the previous one}
    {break;}
}
}

\KwOut{Adaptively task offloading strategy and resource allocation solution.}
\end{algorithm}
\section{Simulation Setup and Results Analysis}
\subsection{Simulation Setup}
\subsubsection{\systemname{} System Settings}
We consider the \systemname{} system, which comprises a cluster of one LEO satellite cluster, $5$ UAVs, 2 BSs, and $10$ mobile users randomly distributed in remote areas. 
To keep the complexity of the simulations tractable while considering a significantly loaded system, the satellite is positioned at an altitude of $780\mbox{ km}$, with the UAVs operating at an altitude of $100 \mbox{ m}$. 
The UAVs fly along a circular trajectory centered at $(1000, 0, 100)\mbox{ m}$ and with a radius of $1000\mbox{ m}$. 
All UAVs are uniformly distributed in the on the flight trajectory and maintain a fly speed of $16.67\mbox{ m/s}$. 
The BSs are assumed to be fixed at the location $(500,0,0)\mbox{ m}$ and $(1500,0, 0)\mbox{ m}$, respectively.  
Based on the communication model described in Section III.C, it is necessary to carefully configure the simulation parameters. 
We consider the communication between UAVs and ground units, utilizing a carrier frequency of $4\mbox{ GHz}$ and a bandwidth of $400\mbox{ MHz}$. 
Following \cite{chen2021energy}, we set the noise spectral density $N_0$ as $-174\mbox{ dBm/Hz}$.
Additionally, the Rician factor $F$ is established at $7$, and the fixed antenna gain $G_0$ at $43.3\mbox{ dBi}$.
For simplicity, each task is assumed to have an average size of $10\mbox{ MB}$. 
All UAVs operate with the same maximum CPU frequency, designated as $f_{m,\max}^{u}=f_{\max}^u=3\times 10^8\mbox{ Hz}$.
The CPU frequency of the server deployed on the satellite is set at $10\times 10^9\mbox{ Hz}$. 
Furthermore, both BSs are equipped with a CPU frequency of $5\times 10^9 \mbox{ Hz}$.
The transmission power for devices communicating with the satellite is set at $5\mbox{ dBm}$. 
For communications from UAVs to the base stations, the transmission power is set at $1.6\mbox{ dBm}$, and from UAVs to the satellite, it is also established at $5\mbox{ dBm}$.

\subsubsection{Algorithm Parameter Settings}
The setup of the DDPG algorithm is defined by a particular group of hyperparameters, outlined as follows: The learning rates for the actor and critic networks are uniformly established at $0.001$. This rate strikes a balance, promoting steady learning while maintaining the stability of the updates.
The discount factor is set at $0.99$, underscoring the emphasis on future rewards in the optimization process. 
A soft update coefficient for target networks is determined to be $0.005$, promoting a smooth transition of weights from the learning networks to the target networks, aiding in the learning process's stability. 
The size of the replay buffer is chosen as $10,000$, offering a vast pool of previous interactions for training, thereby broadening the range of learning instances and mitigating the potential for overfitting recent data. Both the Actor and Critic networks are structured with two fully connected layers.

In the implementation of the SGHS algorithm in our paper,  tailored a set of hyperparameters to enhance the optimization effectiveness. 
The HMS is determined to be $30$.
The number of iterations, which defines the termination condition of the algorithm, is configured to be $10,000$.
The maximum and minimum distance bandwidth, $\mbox{BW}_{\max}$ and $\mbox{BW}_{\min}$ are set as $0.5$ and $5\times 10^{-4}$, respectively. 
The mean values for HMCR and PAR, denoted as $\mu_{\mbox{HMCR}}$ and $\mu_{\mbox{PAR}}$, are established at $0.95$ and $0.3$, respectively. 
Concurrently, the variances for HMCR and PAR, represented as $\sigma_{\mbox{HMCR}}$ and $\sigma_{\mbox{PAR}}$, are set at $0.01$ and $0.05$, respectively.
Additionally, the number of new solutions to be generated is set at $20$.


\subsubsection{Comparison Baselines}
In order to gain insight into the proposed \methodname{}, we select the following four baseline methods for comparison.
\begin{itemize}
\item \textbf{\textit{Random Method}}: both computation offloading and resource allocation are randomly determined.
\item \textbf{\textit{Complete Offloading Method}}: In our proposed \methodname{}, the offloading model is adapted to binary offloading, where tasks are treated as indivisible units that are entirely processed either by a UAV, a nearby BS, or a satellite.
\item \textbf{\textit{Perception Free Method}}: without using the mmWave radar and visual sensors.   
\item \textbf{\textit{Simulated Annealing Algorithm}}: a heuristic optimization algorithm used for allocating computing resources at BSs.
\end{itemize}
\begin{figure}[!t]
\centering
\subfigure[Convergence of DDPG algorithm for problem {\bf P1}.]{
\begin{minipage}[t]{1\linewidth}
\includegraphics[width=3.3in]{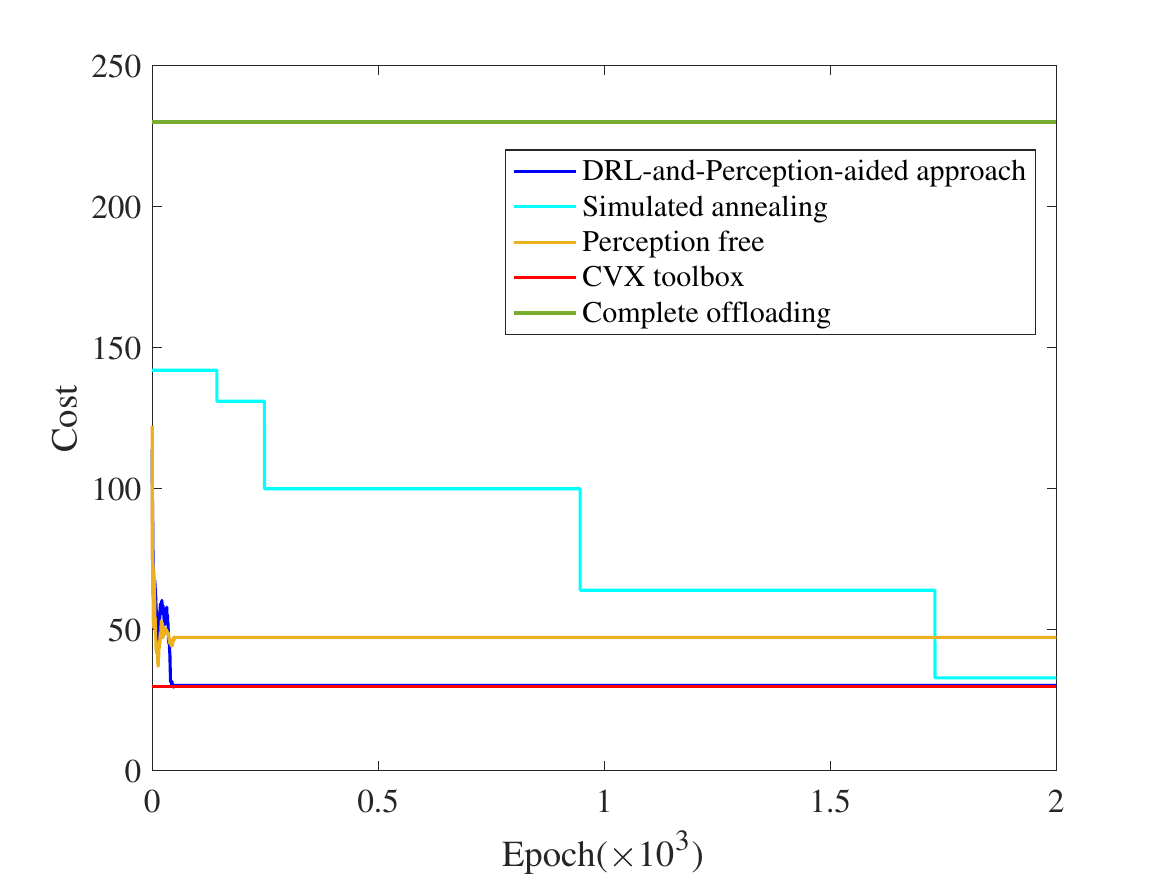}
\end{minipage}
\label{fig:3-1}
}
\subfigure[Convergence of the SGHS algorithm.]{
\centering
\begin{minipage}[t]{1\linewidth}
\includegraphics[width=3.3in]{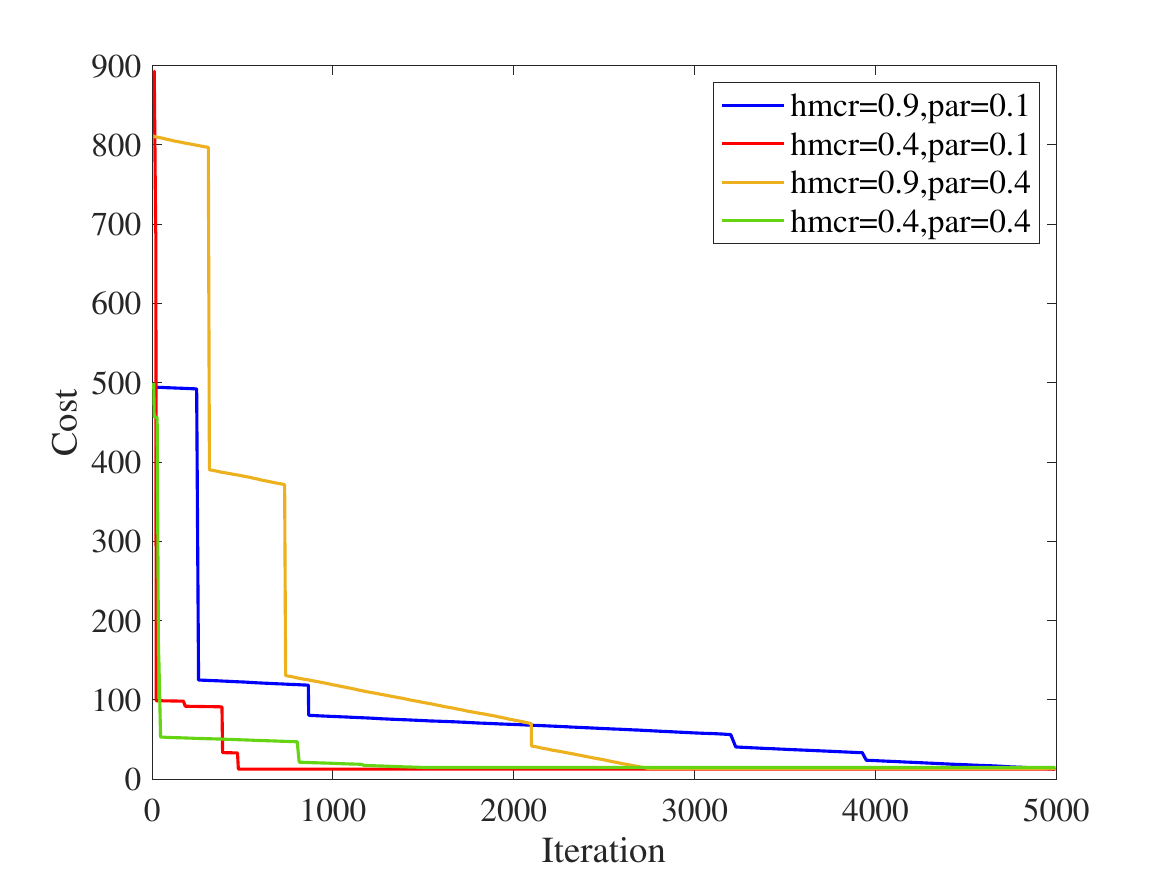}
\end{minipage}
\label{fig:3-2}
}
\caption{Effectiveness of the proposed algorithms for subproblems.}
\end{figure}

\begin{figure}
\centering
\includegraphics[width=3.3in]{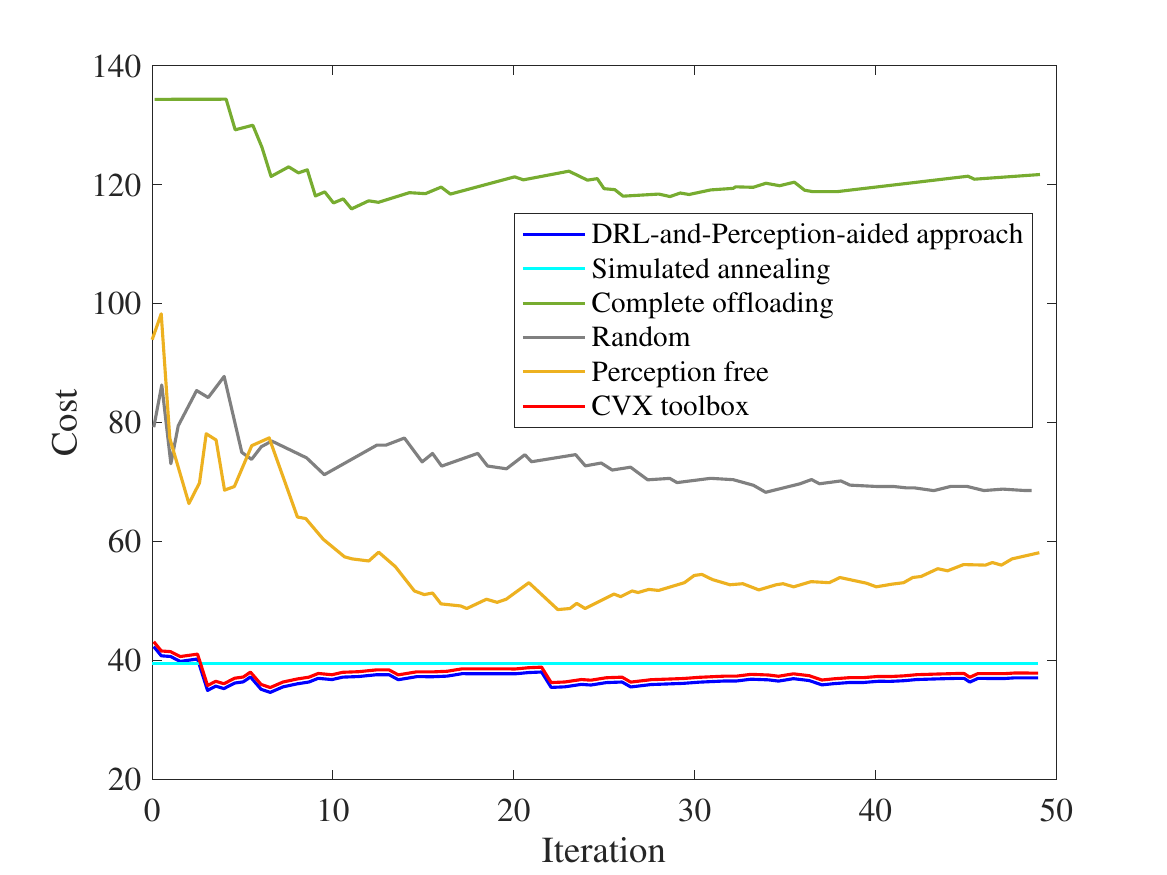}
    \caption{Convergence of the overall optimization process combining three subproblems ({\bf P1, P2, P3}), i.e., Algorithm \ref{algorithm:2}.}
    \label{fig:3-3}
\end{figure}
\begin{figure}[!t]
    \centering
    \includegraphics[width=3.3in]{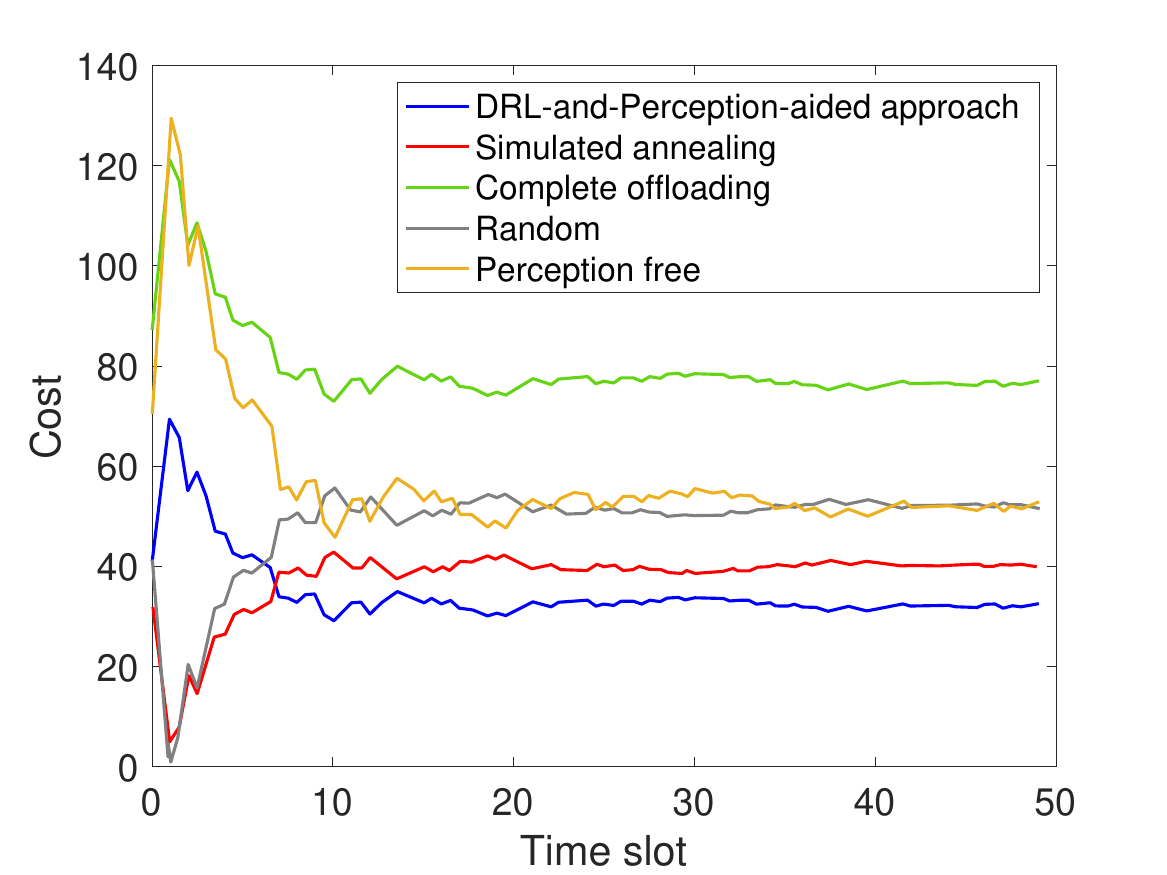}
    \caption{Performance of time-averaged network cost.}
    \label{fig:4}
\end{figure}
\subsection{DRL-and-Perception-aided Approach Verification}
\subsubsection{Convergence Verification of DDPG}
Fig. \ref{fig:3-1} illustrates the convergence and effectiveness performance of the DDPG algorithm for problem {\bf P1}.
From the results, we observe that our proposed \methodname{} converges rapidly to the solution provided by the {\em CVX Toolbox}, which represents the optimal solution for problem {\bf P1}.
The cost performance gap between the \methodname{} and both the Perception Free Method and Complete Offloading Method is more significant than that observed with the Simulated Annealing Algorithm.
By contrast, the Simulated Annealing Algorithm provides a close-to-optimal performance as the \methodname{}, albeit at the expense of significantly increased processing time.
Overall, \methodname{} demonstrates superior performance over the other methods in terms of both reduced cost and shorter processing time.
\begin{figure}[!t]
    \centering
    \includegraphics[width=3.4in]{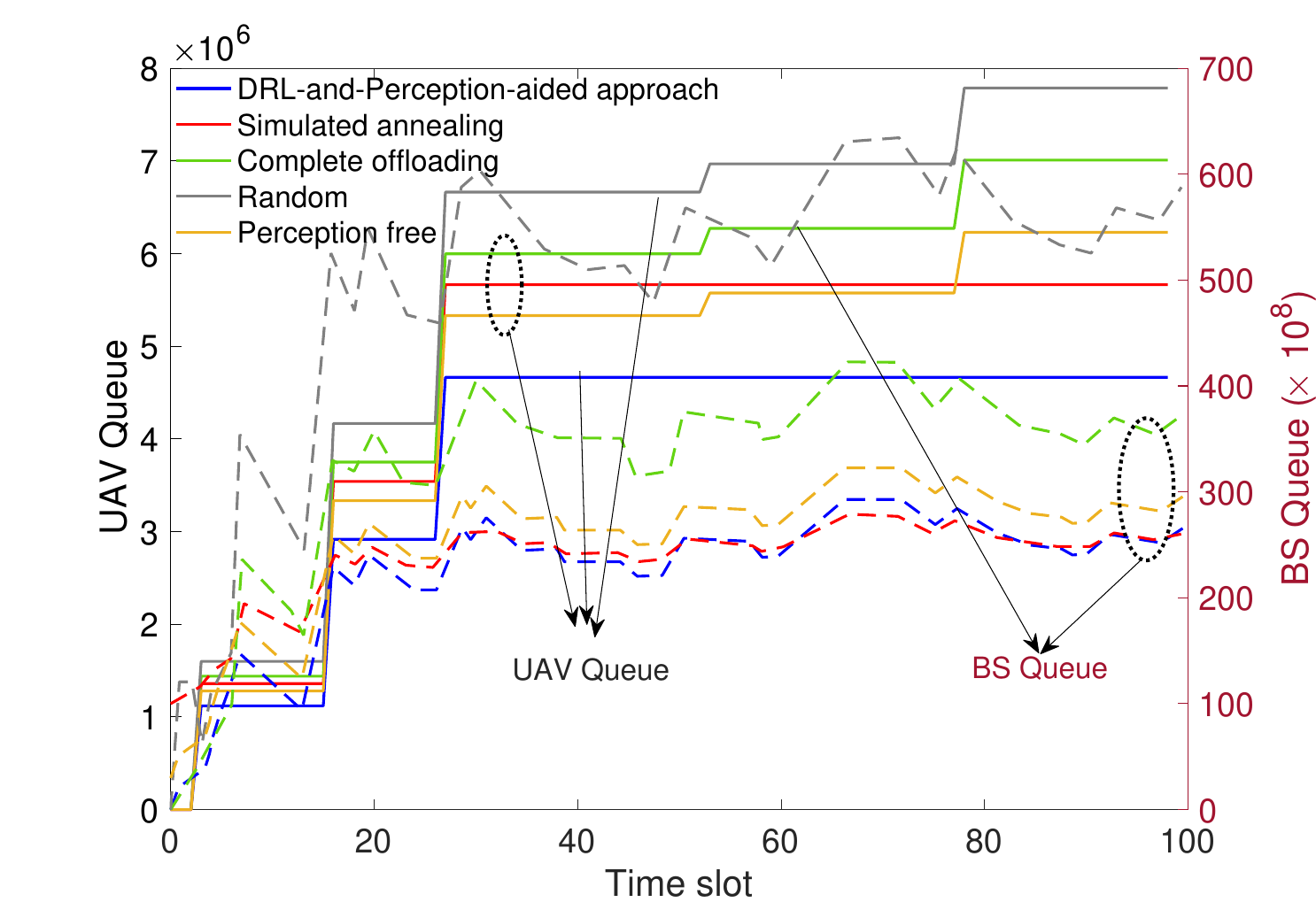}
    \caption{Performance of time-averaged queue backlogs at UAVs and BSs.}
    \label{fig:5}
\end{figure}
\subsubsection{Verification of Effectiveness and Stability of SGHS}
We identify the characteristics of the SGHS (Alg. \ref{algorithm:1}) with respect to parameters $\mbox{HMCR}$ and $\mbox{PAR}$. 
For the given parameter $V=1$ in problem {\bf P3}, Fig. \ref{fig:3-2} illustrates how the values of $\mbox{HMCR}$ and $\mbox{PAR}$ impact the convergence and effectiveness of SGHS.
Accordingly, we conducted four simulation scenarios with combinations of the following values: $\mbox{HMCR}\in\{0.4, 0.9\}$ and $\mbox{PAR}\in\{0.1, 0.4\}.$
Fig. \ref{fig:3-2} demonstrates that a higher HMCR value, notably at $0.9$, leads to a rapid reduction in costs, highlighting its effectiveness in utilizing existing harmonies in the memory to address this particular problem. Conversely, a lower HMCR of $0.4$ shows a more gradual decrease in costs, suggesting that the generation of new random harmonies may not be as advantageous. Additionally, the influence of increasing the PAR from $0.1$ to $0.4$ varies with the HMCR setting; with a high HMCR of $0.9$, an elevated PAR initiates a more aggressive search strategy, yet it converges to a similar endpoint as that observed with a lower PAR. In contrast, at a lower HMCR of $0.4$, a rise in PAR slightly hinders the convergence process, indicating a critical balance between exploring new solutions and exploiting existing ones. Despite these differences in convergence trajectories, all parameter configurations achieve similar cost reductions, underscoring the Harmony Search algorithm's robustness across diverse parameter settings in reaching convergence. This reflects the algorithm’s capacity to consistently find stable solutions, although the path and speed of convergence can vary significantly based on the chosen hyperparameters.

\subsubsection{Effectiveness of The \methodname{}}
In Fig. \ref{fig:3-3}, we evaluate and compare the time-averaged cost performance of the proposed \methodname{} and aforementioned methods under the same mentioned system settings.    
As expected, our proposed \methodname{} outperforms the other baselines significantly. 
By contrast, it is difficult to distinguish the curves for the \methodname{} and the {\em CVX Method}. 
We solve {\bf P1} using {\em CVX} and then address {\bf P2} and {\bf P3} using DQN and SGHS, respectively, under the aforementioned system settings.
Notably, the time-averaged cost associated with \methodname{} is marginally lower than that of the {\em CVX Method}.
This advantage primarily stems from the inherent ability of DRL algorithms to effectively manage time-varying environments.
From the results we notice furthermore that the Simulated Annealing Algorithm towards lower costs slowly and consistently. 
However, it does not achieve the efficiency levels of our proposed approach.

\begin{figure*}[!t]
\centering
\subfigure[]{
\begin{minipage}[t]{0.315\linewidth}
\includegraphics[width=2.6in]{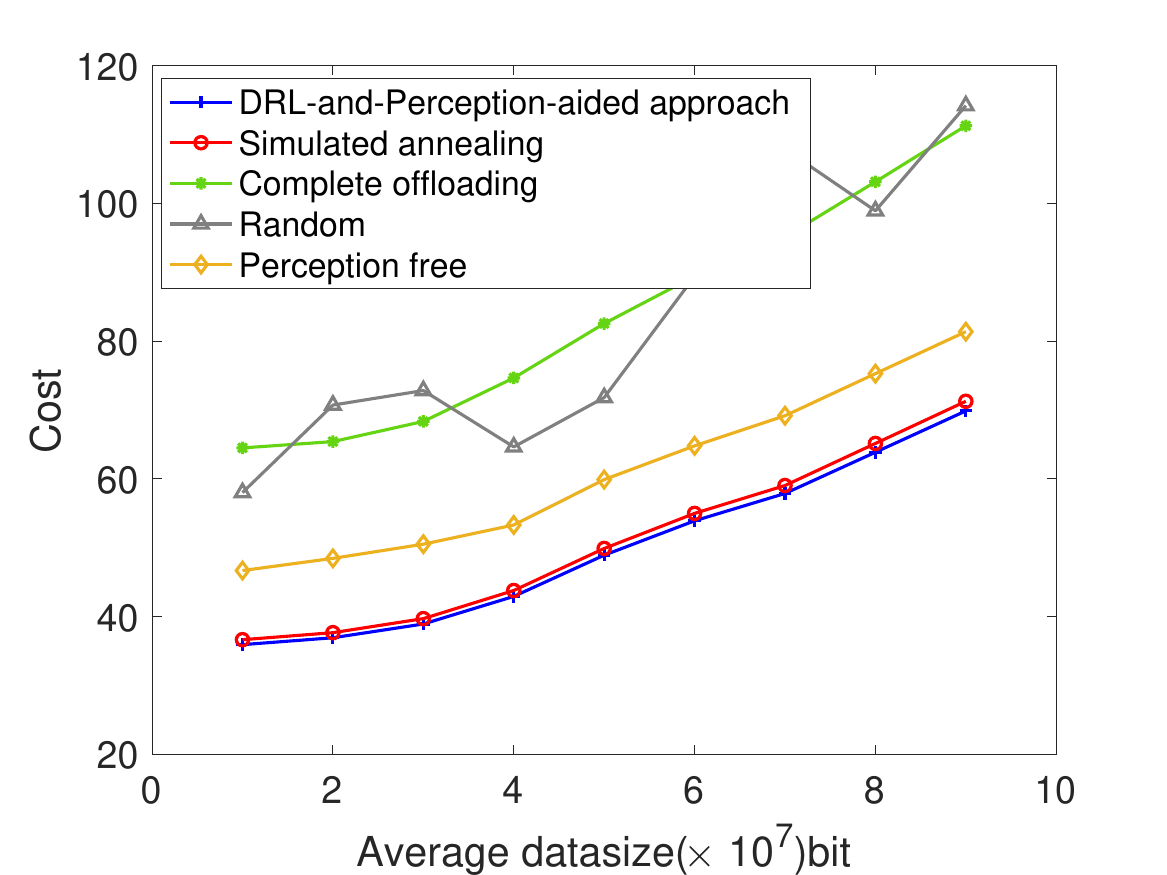}
\end{minipage}
\label{fig:6-1}
}
\subfigure[]{
\begin{minipage}[t]{0.315\linewidth}
\includegraphics[width=2.6in]{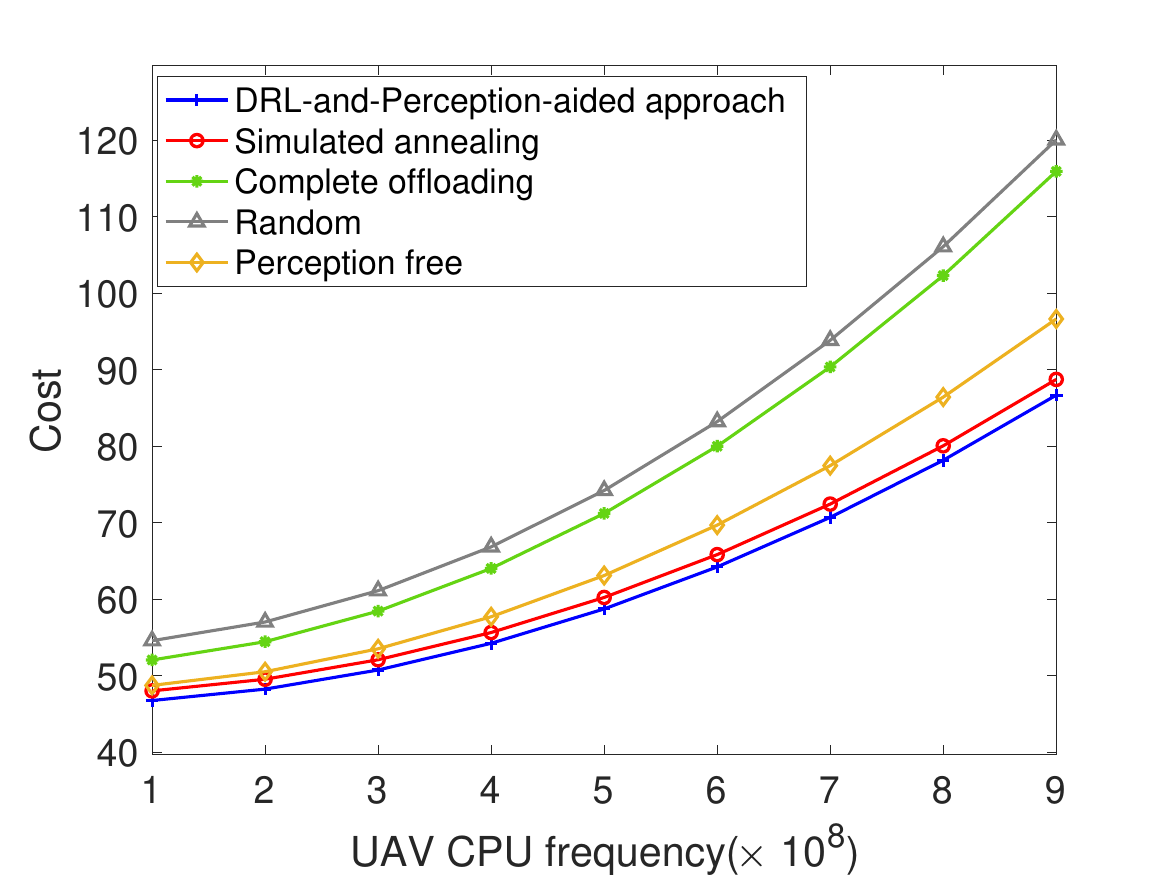}
\end{minipage}
\label{fig:6-2}
}
\subfigure[]{
\begin{minipage}[t]{0.315\linewidth}
\includegraphics[width=2.6in]{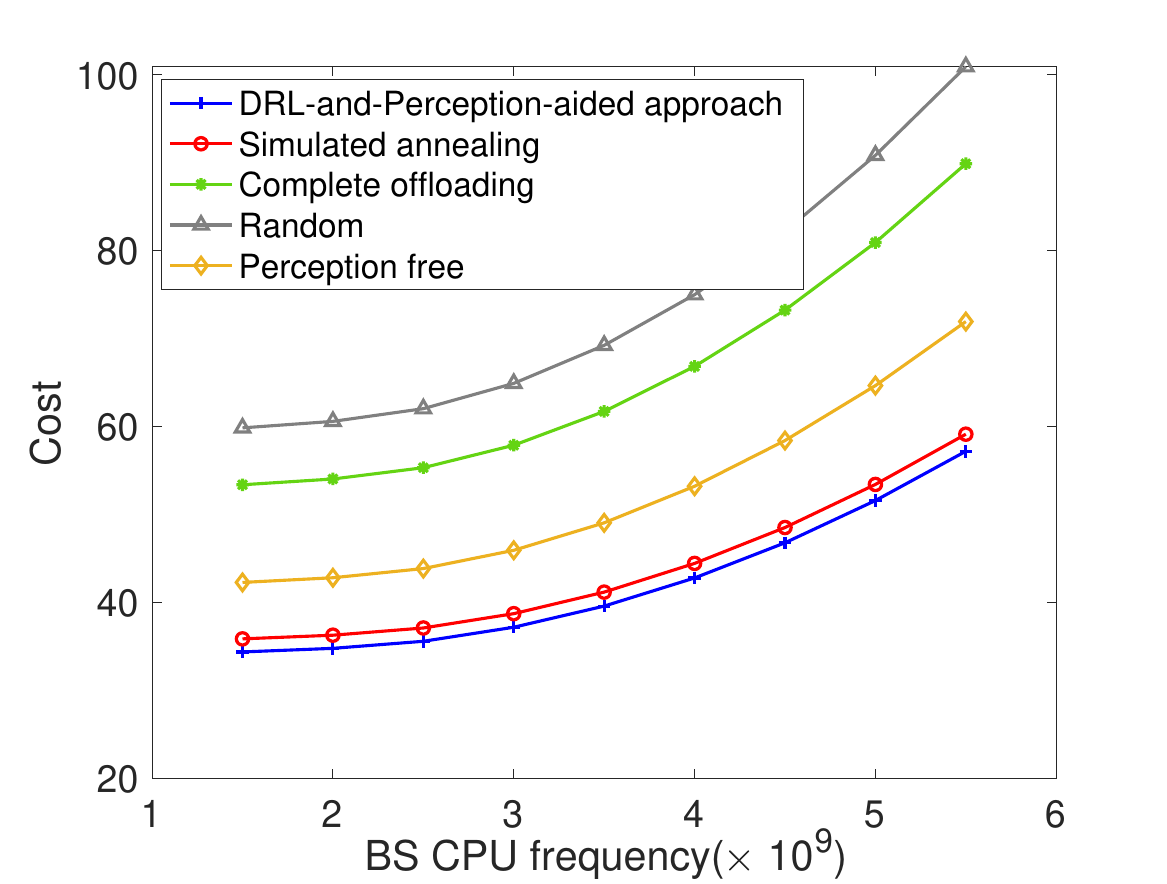}
\end{minipage}
\label{fig:6-3}
}
\caption{Time-averaged cost as a function of  (a) average datasize, (b)  UAV CPU frequency, and (c) BS CPU frequency.}
\end{figure*}
\subsection{Performance Analyse in Dynamic Scenarios}
Fig. \ref{fig:4} illustrates the performance of various algorithms, emphasizing their capability to minimize time-averaged costs amidst the constantly changing conditions of user mobility, location, and task arrival.
We present a detailed classification of user behaviors, categorizing them into a comprehensive set of $60$ distinct types.
The AVA dataset\cite{gu2018ava} focuses on spatiotemporal localization of human actions.
The data is taken from $437$ movies.
These activities range from run/jog, swim, dance to kick, hug, fight.
As seen in Fig. \ref{fig:4}, the proposed \methodname{} demonstrates a rapid and significant reduction in costs, swiftly stabilizing at a low level across various time slots. 
This behavior highlights its rapid adaptability and consistent performance in dynamic environments, characterized by minimal cost fluctuations, underscoring its stability.
In addition to the overall findings in Fig. \ref{fig:4}, the Perception Free Method exhibits the second poorest performance in terms of time-averaged cost.
This underperformance is primarily attributed to its inability to effectively respond to environmental changes.
Regarding the Simulated Annealing Algorithm, although it shows a decreasing trend in costs, it underperforms relative to the proposed \methodname{} because of its slower responsiveness to changes in the environment.

In Fig. \ref{fig:5}, we assess the network stability by examining the UAV average queue $H_m^u(t)$ and the BS average queue $H_{m,n}^{u,b}(t)$ across various time slots.
From the results, we observe that for the \methodname{} on UAV queue, the value of $H_m^u(t)$ asymptotically stabilizes at a relatively low level.
However, both the Random Method and the Complete Offloading Method, due to their failure to account for the limited local computing resources of UAVs and association control, result in poorer performance of queue backlogs $H_m^u(t)$. 
The computation offloading decisions in the Perception Free Method are made without access to perfect state information, which results in larger queue backlogs, denoted as $H_m^u(t)$, compared to those generated by the Simulated Annealing Algorithm. Notably, the queue backlog associated with the proposed \methodname{} progressively stabilizes and approaches a constant value over increasing time slots.
In addition to our observations in Fig. \ref{fig:5} with respect to the UAV queue, similar trends are observed for the BS queue, albeit with distinct nuances. 
It is evident that both the Random Method and the Complete Offloading Method result in poorer performance and higher fluctuations in the BS queue backlog, $H_{m,n}^{bs}(t)$. 
This outcome can be attributed to the fact that the Random Method indiscriminately offloads tasks with equal probability, while the Complete Offloading Method allocates tasks in their entirety without considering the computational capacities of the UAVs and BSs.
By contrast, as illustrated in Fig. \ref{fig:5}, the performance curves for the BS queue of the proposed \methodname{} and the Simulated Annealing Algorithm become indistinguishable when the time slot exceeds $80$.
The Perception Free Method fails to obtain dynamic environment information, resulting in a queue backlog, $H_{m,n}^{bs}$, that is not as short as those observed with the \methodname{} and the Simulated Annealing Algorithm.

\subsection{Performance Demonstration of The Proposed Approach}
\subsubsection{Performance Comparison Versus Datasize}
Fig. \ref{fig:6-1} shows the time-averaged cost achieved by all above methods versus the average datasize. 
From the results, we observe that the average cost achieved by the proposed \methodname{} and Simulated Annealing Algorithm increases as the datasize increases, and outperform the other three methods. 
It is attributed to the proposed algorithm more effective data handling and resource management strategies, which scale more adeptly with increasing loads compared to alternative methods. 
Its consistently lower costs across all data size increments signify a robust methodology that optimizes underlying processes more effectively than other methods, such as Simulated Annealing, Complete Offloading, Random, and Perception Free. 
These latter methods exhibit greater sensitivity to rising data volumes, as evidenced by their steeper cost curves.

\subsubsection{Performance Comparison Versus UAV/BS CPU Frequency}
In Figs. \ref{fig:6-2} and \ref{fig:6-3}, we evaluate and compare the average costs incurred by all aforementioned methods against the maximum CPU frequencies of UAVs, $f_{\max}^{u}$, and BSs, $f_{\max}^{bs}$, respectively. 
The results demonstrate that for both UAV and BS CPU frequencies, the average cost associated with each method increases as the CPU frequencies of the UAVs and BSs increase.
This essentially reflects the fact that increases in CPU frequency, indicative of enhanced processing capabilities, result in higher operational costs for each method.
Additionally, it is observed that the proposed \methodname{} consistently outperforms the other methods, with the performance gap widening as the maximum CPU frequencies of the UAVs and BSs increase. 
This superior performance is primarily due to \methodname{}'s ability to optimally design dynamic task assignments, compute resource allocations, and association controls based on multi-source data, unlike the other methods.

\section{Conclusion}
This paper underscored the significant role of the \systemname{} in the development of 6G telecommunications, especially in enhancing connectivity in remote areas. 
We emphasized the crucial function of UAVs within \systemname{}, equipped with mmWave radar and vision sensors, for reducing uncertainty and improving decision-making through multi-source data collection.
Our paper introduced the \methodname{} designed to jointly optimize task hosting between ground mobile devices and UAVs, computation offloading, association control between UAVs and BSs, and computing resource allocation in real-time within \systemname{}.
This approach focuses on minimizing the time-averaged network cost while maintaining queue stability.
Demonstrated through extensive simulations, our proposed approach outperforms existing benchmarks, highlighting its efficiency and potential for future advancements in 6G network operations. 
This research paves the way for further exploration into optimizing integrated network systems, contributing significantly to the evolution of telecommunications infrastructure.

\bibliography{reference}
\bibliographystyle{IEEEtran}

\end{document}